\documentclass[12pt]{article}
\usepackage{amsmath,amsfonts,amscd,theorem,color}

\definecolor{airforceblue}{rgb}{0.36,0.54,0.66}
\definecolor{amaranth}{rgb}{0.9,0.17,0.31}
\definecolor{amber}{rgb}{1.0,0.75,0.0}
\definecolor{ao}{rgb}{0.0,0.0,1.0}
\definecolor{aoenglish}{rgb}{0.0,0.5,0.0}
\definecolor{applegreen}{rgb}{0.55,0.71,0.0}
\definecolor{armygreen}{rgb}{0.29,0.33,0.13}
\definecolor{asparagus}{rgb}{0.53,0.66,0.42}
\definecolor{auburn}{rgb}{0.43,0.21,0.1}

\newcommand{\za}{\alpha}

\newcommand{\zg}{\gamma}

\begin{document}

\title{On two possible ways to recover Ordinary Thermodynamics from Extended Thermodynamics of Polyatomic gases}
\author{F. Demontis, S. Pennisi \\
Department of di Mathematics and Computer Science, University of Cagliari, \\
 09124 Cagliari, Italy;
fdemontis@unica.it; spennisi@unica.it}
\date{}
\maketitle
 \small {\em \noindent }
\hspace{-2.5 cm}

\abstract{ We consider two possible ways, i.e., the Maxwellian Iteration and the Chapman-Enskog Method, to recover Relativistic Ordinary Thermodynamics from Relativistic Extended Thermodynamics of Polyatomic gases with $N$ moments. Both of these methods give the Eckart equations which are the relativistic version of the Navier-Stokes and Fourier laws as a first iteration. However, these methods do not lead to the same expressions of the heat conductivity $\chi$, the shear viscosity $\mu$, and the bulk viscosity $\nu$ which appear as coefficients in the Eckart equations. In particular, we prove that the expressions of $\chi$, $\mu$, and $\nu$ obtained via the Chapman-Enskog method do not depend on $N$, while those obtained through the Maxwellian Iteration depend on $N$. Moreover, we also prove that these two methods lead to the same results in the nonrelativistic limit}.

\section{Introduction}
Rational Extended Thermodynamics (RET) is an elegant theory appreciated by mathematicians and physicist. This theory 
was developed in a systematic way by Liu and M\"uller in \cite{MUS} for the classical case, while the relativistic case was considered by Liu, M\"uller and Ruggeri in \cite{MUr}. Both  articles \cite{MUS} and \cite{MUr} are based on few natural assumptions, in fact only universal principles\footnote{The universal principles used to develop the RET are: 1) the entropy principle, 2) the causality principle and 3) the galileian principle in the classical case or the relativity principle in the relativistic case.} are imposed and, as a consequence of these principles, the hyperbolicity of the field equations is established. This is an important achievement because in this way in the relativistic case the paradox of infinity velocity of the propagating waves is automatically eliminated. These and other related results can also be found in the book \cite{M-R} but they concern only the case of a monoatomic gas. The extension to the case of polyatomic gases was done in \cite{A.T-R-S} for the classical case and in \cite{P-R} for the relativistic case. More details on these generalizations to the polyatomic case can also be found in the book \cite{R-S}. However, a physical observation by Pennisi recently published in \cite{Pen} has caused a revision of the previous models both for the classical and relativistic cases leading to new results that can be found in \cite{1A-C-P-R} and \cite{CPTR1}.
In particular, in the  article \cite{CPTR1} a satisfactory model for the Relativistic Extended Thermodynamics of Polyatomic gases with $N$ moments ($ET^N$) has been proposed.

However, so far, Ordinary Thermodynamics (OT) has been succesfully used in practical applications, so a good
test for establishing the validity of RET consists of finding procedures of approximation
which allow us to get the equations of OT as a first step. In the literature two procedures have been proposed which realize this objective: the Maxwellian Iteration (MI) and the Chapman-Enskog Method (CEM). \newline
We will prove that the application of both in the relativistic case of these procedures
leads, as a first iteration, to the Eckart equations \cite{Eck}, which in \cite{MUr} are called the relativistic version of the Navier-Stokes and Fourier laws and are two fundamental laws of Relativistic Ordinary Thermodynamics (ROT). From now on we refer to the Eckart equations as the Navier-Stokes and Fourier laws.  It is important to remark that in the Navier-Stokes and Fourier equations the following important quantities
appear as coefficients:  the heat conductivity $\chi$, the shear viscosity $\mu$, and the bulk viscosity $\nu$. The aim of this paper is to show that the expressions of $\chi$, $\mu$, and $\nu$ obtained via the Chapman-Enskog method do not depend on $N$, whereas these expressions obtained through the Maxwellian Iteration depend on $N$. In order to get this result, let us recall the basic facts on the field equation of Relativistic Extended Thermodynamics of Polyatomic gases and Relativistic Ordinary Thermodynamics.\newline  
Let us start by considering the balance equations of ET$^N$ \cite{CPTR1}. They are obtained starting from the Boltzmann equation 
\begin{align}\label{1}
\begin{split}
& p^\alpha \, \partial_\alpha \, f = Q \, , \mbox{with} \quad Q= \frac{U^\mu p_\mu}{c^2 \tau}\left[f_E-f-f_E \, p^\gamma q_\gamma \frac{3}{m \, c^4 \rho \, \theta_{1,2}} \, \left(1+\frac{\mathcal{I}}{m c^2}\right) \right] \, , \\
& f= e^{-1 \, - \, \frac{\chi}{k_B}} \, , \quad
\chi= \sum_{n=0}^{N}   \frac{1}{m^{n-1}} \lambda_{\alpha_1 \cdots \alpha_n} p^{\alpha_1}  \cdots p^{\alpha_n}\left( 1 \, + \, \frac{\mathcal{I}}{m \, c^2} \right)^n \, , 
\\
& f_E = e^{-1 \, - \, \frac{m \, \lambda_E \, + \, \frac{U_\mu}{T} \, p^\mu \, \left( 1 \, + \, \frac{\mathcal{I}}{m \, c^2} \right)}{k_B}} \, ,  \quad  \theta_{1,2} = \frac{3 \, p}{\rho^2 \, c^4}  \,
\left( e \, + \, p  \right)    \,  . 
\end{split}
\end{align}
Here, $f$ is the distribution function, $k_B$ the Boltzmann constant, $m$ the relativistic particle mass, $\lambda_{\alpha_1 \cdots \alpha_n}$ are Lagrange multipliers, $c$ the light speed,  $p^\mu$ the 4-momentum of the particle (from now on the greek indexes take the values $0, 1, 2,3$) such that $p_\alpha  p^\alpha =m^2 c^2$, $\mathcal{I}$ is the internal energy of the particle due to rotational and vibrational modes, $\tau$ a relaxation time, $\lambda_E$ the first Lagrange multiplier calculated at equilibrium, $U^\mu$ the 4-velocity  such that $U_\alpha  U^\alpha =c^2$, $Q$ the production term in the Boltzmann equation, $T$ the absolute temperature, $\rho$
the mass density, $p$ the pressure, $e$ the energy, and $q^\alpha$ the heat flux such that  $U_\alpha  q^\alpha =0$. The name ``Lagrange multipliers'' has been used in literature \cite{MUS, MUr, M-R} and this terminology is due to the fact that the distribution function $f$ can be obtained through a variational principle called the Maximum Entropy Principle (MEP) \cite{P-R} with constrained variables. In the case with six moments, the heat flux is zero and $q^\alpha$ replaces the scalar anonimous quantities which are present in \cite{CK} and $q^\alpha$ is an unknown function to be determined by imposing that the production of mass and energy-momentum are zero. \\
After that, the function $\varphi(\mathcal{I})$ is introduced which measures "how much" the gas is polyatomic.\newline
Finally, by multiplying \eqref{1}$_1$ by $\frac{c}{m^{n-1}} \,  p^{\alpha_1}  \cdots p^{\alpha_n} \left( 1 \, + \, \frac{\mathcal{I}}{m \, c^2} \right)^n \,  \varphi(\mathcal{I})$ and integrating the result with respect to $d \, \mathcal{I} \, d \, \vec{P}$ one obtains the balance equations
\begin{subequations} 
\begin{align}\label{2_a}
& \partial_{\alpha} A^{\alpha} = 0 \, , \quad 
\partial_{\alpha} A^{\alpha \alpha_1} = 0 \, ,  \\ \nonumber
& \partial_{\alpha} A^{\alpha \alpha_1 \cdots \alpha_n} = I^{\alpha_1 \cdots \alpha_n}  \, , \quad \mbox{for} \quad n= 2 , \, \cdots \, N \,  ,
\end{align}
where 
\begin{align}\label{2_b}
& A^{\alpha_1 \cdots \alpha_{n+1}}= \frac{c}{m^{n-1}} \int_{\Re^3} \int_{0}^{+ \infty} f \, p^{\alpha_1} \cdots p^{\alpha_{n+1}} \left( 1 \, + \, \frac{\mathcal{I}}{m \, c^2} \right)^n \,  \varphi(\mathcal{I}) \, d \, \mathcal{I} \, d \, \vec{P} \, , \\ \nonumber
& I^{\alpha_1 \cdots \alpha_{n}}= \frac{c}{m^{n-1}} \int_{\Re^3} \int_{0}^{+ \infty} Q \, p^{\alpha_1} \cdots p^{\alpha_n} \left( 1 \, + \, \frac{\mathcal{I}}{m \, c^2} \right)^n \,  \varphi(\mathcal{I}) \, d \, \mathcal{I} \, d \, \vec{P}  \, . 
\end{align}
\end{subequations}
Obviously, eqs. \eqref{2_a}$_{1,2}$ are particular cases of \eqref{2_a}$_{3}$ with $n=0,1$ but it is better to write them separately because they are the mass and energy-momentum conservation laws, respectively; their productions $I$ and $I^{\alpha_1}$ are zero as a consequence of the definition of $Q$ (see \eqref{1}$_{2}$). Moreover, we sometimes denote $A^{\alpha}$ with $V^{\za}$ and $A^{\alpha \alpha_1}$ with 
$T^{\alpha \alpha_1}$. \newline
Then, the Lagrange multipliers $\lambda_{\alpha_1 \cdots \alpha_n}$ are obtained in terms of the physical variables but in a linear departure from equilibrium (here denoted with the suffix $E$) which is defined as the status where $\lambda_{\alpha_1 \cdots \alpha_n}=0$ for $n=2, \, \cdots \, N$ and $\lambda_{\alpha}^E= \frac{U_\alpha}{T}$. The calculations for the case $N=2$ can be found in \cite{CPTR1}. These calculations are based on the expression of the following tensor 
\begin{align}\label{3}
A^{\alpha_1 \cdots \alpha_{n+1}}_E = \frac{c}{m^{n-1}} \int_{\Re^3} \int_{0}^{+ \, \infty} f_E \, p^{\alpha_1} \, \cdots \, p^{\alpha_{n+1}} \, \left( 1 + \, \frac{\mathcal{I}}{m \, c^2} \right)^n \, \varphi(\mathcal{I}) \, d \, \mathcal{I}  \, d \, \vec{P} \, . 
\end{align}
This tensor is only determined in terms of the energy $e$ which is given by 
\begin{align}\label{4}
\frac{e}{\rho \, c^2} = \frac{\int_{0}^{+ \infty} J^*_{2,2} \, \left( 1 + \, \frac{\mathcal{I}}{m \, c^2} \right) \, \varphi(\mathcal{I}) \, d \, \mathcal{I}}{\int_{0}^{+ \infty} J^*_{2,1} \, \varphi(\mathcal{I}) \, d \, \mathcal{I}} \, ,
\end{align}
where $J_{m,n}(\gamma)= \int_0^{\infty} e^{-\zg \cosh{s}}\cosh^n{s}\sinh^m{s}\, ds, \,\, \zg=\frac{m c^2}{k_B T}, \,\, J^*_{m,n}= J_{m,n}\left[\zg\left(1+\frac{\mathcal{I}}{mc^2}\right)\right]\,.$
Here we report a new short proof of this result because we need to use $A^{\alpha_1 \cdots \alpha_{n+1}}_E$. In fact, from \eqref{3} it follows that 
\begin{align}\label{5}
d \, A^{\alpha_1 \cdots \alpha_{n+1}}_E = - \, \frac{m}{k_B} \,  \left( A^{\alpha_1 \cdots \alpha_{n+1}}_E d \, \lambda^E \, + \, A^{\alpha_1 \cdots \alpha_{n+2}}_E d \, \lambda^E_{\alpha_{n+2}} \right) \, . 
\end{align}
This equation, written for $n=0$, is 
\begin{align*}
d \, \left( \rho \, U^{\alpha_1} \right) = - \, \frac{m}{k_B} \,  \left[ \rho \, U^{\alpha_1} \,  d \, \lambda^E \, + \, 
\left( e \, \frac{U^{\alpha_1} U^{\alpha_2} }{c^2} \, + \, p \, 
h^{\alpha_1 \alpha_{2}} \right) d \, \lambda^E_{\alpha_{2}} \right] \, , 
\end{align*}
whose contraction with $U^{\alpha_1}$ allows us to determine 
\begin{align*}
d \, \lambda^E = - \, \frac{k_B}{m \, \rho} \, d \, \rho \, - \, \frac{e}{\rho \, c^2}
\, U^{\alpha_2} \,  d \, \lambda^E_{\alpha_{2}}  \, .
\end{align*}
By substituting this in eq. \eqref{5}, we find 
\begin{align*}
d \, A^{\alpha_1 \cdots \alpha_{n+1}}_E = A^{\alpha_1 \cdots \alpha_{n+1}}_E  \,  \left( \frac{1}{\rho} \, d \, \rho \, + \, \frac{e \, m}{\rho \, c^2 k_B}
\, U^\gamma \,  d \, \lambda^E_\gamma \right) \, - \, \frac{m}{k_B} \, A^{\alpha_1 \cdots \alpha_{n+2}}_E d \, \lambda^E_{\alpha_{n+2}}  \, . 
\end{align*}
If we take $\rho$ and $\lambda_\gamma^E$ as independent variables, the coefficient of $d \, \rho $ shows that $A^{\alpha_1 \cdots \alpha_{n+1}}_E$ is linear and homogeneous in the variable $\rho $, while the coefficient of $d \, \lambda^E_\gamma $ allows us to determine 
\begin{align}\label{6}
A^{\alpha_1 \cdots \alpha_{n+2}}_E =- \, \frac{k_B}{m} \, \frac{\partial \, A^{\alpha_1 \cdots \alpha_{n+1}}_E}{\partial \, \lambda^E_{\alpha_{n+2}} }\, + \, \frac{e}{\rho \, c^2}
\,  A^{\alpha_1 \cdots \alpha_{n+1}}_E \, U^{\alpha_{n+2}}  \, . 
\end{align}
Taking into account this result, all the tensors $A^{\alpha_1 \cdots \alpha_{n+1}}_E$  are determined in terms of the previous ones. Obviously, we must  be careful and express everything in terms of  $\rho$ and $\lambda_\gamma^E$. Regarding $\lambda_\gamma^E$, we note that 
\begin{align*}
\lambda_\gamma^E = \frac{U_\gamma}{T} \quad \rightarrow \quad T = \frac{c}{\sqrt{\lambda_\delta^E \lambda^{E \delta}}} \, ; \quad U_\gamma = \frac{c}{\sqrt{\lambda_\delta^E \lambda^{E \delta}}} \, \lambda_\gamma^E \, ; \quad A^\gamma_E = \rho \, U^\gamma = \frac{\rho \, c}{\sqrt{\lambda_\delta^E \lambda^{E \delta}}} \, \lambda_\gamma^E \, . 
\end{align*}
As a test, let us consider eq. \eqref{6} for $n=0$, i.e., 
\begin{align*}
T^{\alpha_1 \alpha_2}_E &=- \, \frac{k_B}{m} \, \frac{\partial \, A_{E}^{\alpha_1}}{\partial \, \lambda^E_{\alpha_2} }\, + \, \frac{e}{\rho \, c^2}
\,  A^{\alpha_1}_E \, U^{\alpha_{2}} \\ 
&= - \, \frac{k_B}{m} \left( \frac{\rho \, c}{\sqrt{\lambda_\delta^E \lambda^{E \delta}}} g^{\alpha_1 \alpha_2} \, - \, 
\frac{\rho \, c}{\left(\lambda_\delta^E \lambda^{E \delta}\right)^{3/2}} \lambda^{\alpha_1}_E \, \lambda_E^{\alpha_{2}} \right) \, + 
 \,  \frac{e}{\lambda_\delta^E \lambda^{E \delta}} \lambda^{\alpha_1}_E \, \lambda_E^{\alpha_{2}}\\ 
&= \frac{k_B}{m} \, \rho \, T \, h^{\alpha_1 \alpha_2} \, + \, \frac{e}{c^2} \, U^{\alpha_1}  U^{\alpha_{2}}  \, . 
\end{align*}
So we have obtained the expression for the coefficient of $U^{\alpha_1}  U^{\alpha_{2}} $, while the other term gives 
\begin{align}\label{7}
p = n \, k_B \, T   \, .
\end{align}
{\bf Note}: In the above calculations we have used the property 
\begin{align*}
\lambda^{E \gamma} \lambda^E_\gamma= g^{\mu \gamma} \, \lambda^E_\mu \lambda^E_\gamma
\quad \rightarrow \quad \frac{\partial}{\partial \, \lambda^E_\delta} \left( \lambda^{E \gamma} \lambda^E_\gamma \right) = 2 \,  g^{\mu \delta} \, \lambda^E_\mu = 2 \,  \lambda^{E \delta} \, .
\end{align*}
We note also that \eqref{6} does not allow us to obtain the expression for the energy $e$; so to find it we must go back to the definitions \eqref{3} for $n=0$ and $n=1$ and contract them by $U_{\alpha_1}$ and $U_{\alpha_1} U_{\alpha_2}$, respectively. The  results are 
\begin{align*}
\begin{split}
& \rho \, c^2 = \frac{4 \, \pi \, m^4 c^5}{\sqrt{- \, g}} \, e^{-1 - \, \frac{m \, \lambda^E}{k_B}} \int_{0}^{+ \infty} J^*_{2,1} \, \varphi(\mathcal{I}) \, d \, \mathcal{I} \, , \\
& e \, c^2 = \frac{4 \, \pi \, m^4 c^7}{\sqrt{- \, g}} \, e^{-1 - \, \frac{m \, \lambda^E}{k_B}} \int_{0}^{+ \infty} J^*_{2,2} \, \left( 1 + \, \frac{\mathcal{I}}{m \, c^2} \right) \, \varphi(\mathcal{I}) \, d \, \mathcal{I} \, , 
\end{split}
\end{align*}
the second one of these expressions, divided by the first one gives the formula \eqref{4} reported above.\newline
From these considerations it is possible to derive the expression for $A^{\alpha \alpha_1 \cdots \alpha_j  }_E$ reported in eqs. (29) and (30) of \cite{Pen} and rewritten in eqs.(14)-(16) of \cite{CPTR1}. 
For the convenience of the reader we 
write this expression  below:
\begin{align}\label{AE}
A^{\alpha_1 \cdots \alpha_{j+1 } }_E= \sum_{k=0}^{\left[ \frac{j+1}{2} \right]} \rho c^{2k} \theta_{k,j} \, h^{( \alpha_1 \alpha_2 } \cdots h^{ \alpha_{2k-1} \alpha_{2k} } U^{\alpha_{2k+1} } \cdots U^{\alpha_{j+1} ) } \, ,
\end{align}
where the round brackets appearing in $h^{( \alpha_1 \alpha_2 } \cdots h^{ \alpha_{2k-1} \alpha_{2k} } U^{\alpha_{2k+1} }\cdots U^{\alpha_{j+1} ) }$ denote the symmetric part of this tensor, 
while the scalar coefficients $\theta_{k,j}$ are defined as follows:
\begin{align}\label{SC}
\theta_{k,j} = \frac{1}{2k+1}  \begin{pmatrix}
j+1 \\ 2k
\end{pmatrix} \frac{\int_0^{+\infty} J_{2k+2,j+1-2k}^* \, \left( 1 + \frac{\mathcal{I}}{m c^2} \right)^j \, \phi(\mathcal{I})  \, d \, \mathcal{I}}{\int_0^{+\infty} J_{2,1}^* \,  \phi(\mathcal{I})  \, d \, \mathcal{I}} \, .
\end{align}
Moreover, $\theta_{k,j}$ can be determined by the 
recurrence relations which use the quantity $\gamma= \frac{m \, c^2}{k_B T}$:
\begin{align}\label{AE1}
\begin{split}
& \theta_{0,0}=1 \, ,\\
& \theta_{0,j+1} = \frac{e}{\rho \, c^2} \, \theta_{0,j}  \, - \, \frac{\partial \, \theta_{0,j}}{\partial \, \gamma}\, ,   \\
& \\
&  \theta_{h,j+1} = \frac{j+2}{\gamma} \left( \theta_{h,j} + \frac{j+3-2h}{2h} \theta_{h-1,j} \right)\, ,  \hspace{1.9 cm} \mbox{for } h=1, \, \cdots , \,\left[ \frac{j+1}{2} \right] \, , \\
& \\
&  \theta_{\frac{j+2}{2},j+1} = \frac{1}{\gamma} \theta_{\frac{j}{2}, \, j}\, ,  \hspace{6.3 cm} \mbox{for $j$ even}  \, .
\end{split}
\end{align}
Regarding the production terms in the balance equations \eqref{2_b}$_2$ we see that, by means of \eqref{3}, it becomes 
\begin{align}\label{8}
\begin{split}
& I^{\alpha_1 \cdots  \alpha_i} = - \, \frac{U_\alpha}{c^2 \tau} \,  \left( A^{\alpha \alpha_1 \cdots  \alpha_i} \, - \, A^{\alpha \alpha_1 \cdots  \alpha_i}_E \right) \, - \, \frac{3}{\rho \, c^6 \, \tau \, \theta_{1,2} } \, U_\alpha \, q_\beta \,  A^{\alpha \beta \alpha_1 \cdots  \alpha_i}_E \, , \\
& \mbox{where} \quad q^\beta = U_\gamma \, \left( T^{\gamma \beta} \, - \, T^{\gamma \beta}_E \right) \, .
\end{split}
\end{align}
So far we have described the results of Extended Thermodynamics of polyatomic gases, as obtained in \cite{CPTR1} following the new ideas of \cite{Pen}. \\
On the other hand, {\bf Relativistic Ordinary Thermodynamics} (ROT) uses only the equations \eqref{2_a}
with the following definitions
\begin{align}\label{9}
\begin{split}
& A^{\alpha \alpha_1} = T^{\alpha \alpha_1} = \frac{e }{c^2} \, U^\alpha  U^{\alpha_1} \, + \, (p+ \pi) h^{\alpha \alpha_1} \, + \, \frac{2 }{c^2} \, U^{( \alpha} q^{\alpha_1 )}  \, + \, t^{< \alpha \alpha_1 > } \quad \mbox{where} \\
& \pi = - \nu \, \partial_{\alpha} \, U^\alpha \, , \, 
q^\beta = - \, \chi \, h^{\alpha \beta}
\left( \partial_{\alpha} \, T \, - \, \frac{T}{c^2} \, U^\mu \, \partial_\mu \, U_\alpha \right) \, , \,
t_{< \beta \gamma >} = 2 \, \mu \, h^\alpha_\beta \, h^\mu_\gamma \, \partial_{< \alpha} \, U_{\mu >} \, ,
\end{split}
\end{align}
where the angular brackets appearing in $t_{< \beta \gamma >}$ denote the traceless symmetric part of the tensor.\newline 
Here \eqref{9}$_{2-4}$ are the Eckart equations \cite{Eck}; in particular, following \cite{MUr} equation \eqref{9}$_{3}$ corresponds to relativistic version of the Fourier law 
while equations \eqref{9}$_{2}$ and \eqref{9}$_{4}$ correspond to the relativistic version of the Navier-Stokes law. 
The coefficients $\nu$, $\chi$, $\mu$ are called the bulk viscosity, the heat conductivity and the shear viscosity, respectively. \newline
These equations have the drawback that they are not hyperbolic but parabolic. As we have already said, this was the reason for the birth of Extended Thermodynamics whose equations are hyperbolic. 

The paper is organized as follows: In section \ref{sec2} we briefly recall how the MI procedures works and show how it is possible to reconstruct the laws of Relativistic Ordinary Thermodynamics by using this procedure. 
Moreover, in section \ref{sec2} we derive the expressions of the heat conductivity $\chi$, the shear viscosity $\mu$, and the bulk viscosity $\nu$ in the particular cases $N=3$ and $N=2$ putting in evidence that, if one uses the MI procedure, these expressions depend on the number of moments $N$. In section \ref{sec3} we explain how the CEM procedure works and derive the laws of the ROT by using this procedure. In section \ref{sec:NR}, generalizing a well-known result of \cite{1b} for monoatomic case, we prove that in the non-relativistic case the MI and the CEM procedures lead to the same results. The convergence of $\nu$, $\chi$, $\mu$ in the non relativistic limit is proved in section \ref{sec:NR_1}. Then, the results obtained are summarized in section \ref{sec4}. Finally, an appendix is devoted to some particular integrals used to develop the computations of this paper.

\section{The Maxwellian Iteration}\label{sec2}
The Maxwellian Iteration Method was applied to recover OT from Extended Thermodynamics of monoatomic gases in \cite{MUS}
for the non relativistic case and in \cite{MUr} for the relativistic framework. The relativistic case for polyatomic gases with $N=2$, and for its subsystems with fourteen and six moments, has been treated in \cite{CPTR1}. In the next section MI will be implemented in the case of an arbitrary number $N$. It works in the following way:
\begin{itemize}
	\item The eqs. \eqref{2_a} are considered, but with their left hand sides calculated at equilibrium and their right hand sides at first order with respect to equilibrium, i.e., 
\begin{align}\label{10}
\begin{split}
& \partial_{\alpha} A^{\alpha}_E = 0 \, , \quad 
\partial_{\alpha} A^{\alpha \alpha_1}_E = 0 \, ,  \\
& \partial_{\alpha} A^{\alpha \alpha_1 \cdots \alpha_n}_E = I^{\alpha_1 \cdots \alpha_n}_{MI}  \, , \quad \mbox{for} \quad n= 2 , \, \cdots \, N \,  ,
\end{split}
\end{align}
where the meaning of the subscrpt MI will be introduced in the next item.
\item The deviations of the independent variables from equilibrium are calculated in terms of $\partial_{\alpha} \lambda^E$ and $\partial_{\alpha} \lambda_\mu^E$ from \eqref{10}$_3$; they are called "first iterates" and we will denote them with a suffix $MI$. After that, they are substituted in $T^{\alpha \beta} \, - \, T^{\alpha \beta}_E$ with $T^{\alpha \beta}$ given by \eqref{9}$_1$. 
\item The quantities $\partial_{\alpha} \lambda^E$ and $U^\alpha U^\mu \partial_{\alpha} \lambda_\mu^E$ are calculated from \eqref{10}$_{1,2}$ and substituted in the expression of $T^{\alpha \beta} \, - \, T^{\alpha \beta}_E$ obtained in the previous step. 
\end{itemize}
In this way one obtains \eqref{9} of {R}OT, with particular expressions $\nu_{MI}^N$,  $\chi_{MI}^N$, $\mu_{MI}^N$ of the bulk viscosity $\nu$, the heat conductivity $\chi$ and the shear viscosity $\mu$. \\
However, these expressions depend on the number $N$ of the extended model from which they come from. For example, in \cite{CPTR1} it was found that, for the subsystem with 14 moments, the values of $\mu$ and $\chi$ remain the same as for the model with 15 moments (i.e., $N=2$) while $\nu$ changes. This expression for $\nu$ changes again for its further subsystem
with six moments (In this subsystem with six moments $\mu$ and $\chi$ do not play a role). This is not fully satisfactory because there is only one Ordinary Thermodynamics and it is strange that its equations depend on the number $N$ of the extended model from which they are derived. In the next subsection this Maxwellian iteration will be described in more detail for an arbitrary number $N$; furthermore, we will see what is the difference of the expressions of $\mu$, $\chi$ and $\nu$ in the cases $N=3$ and $N=2$ in the subsections \ref{sec:2_2} and \ref{sec:2_3}, respectively.

\subsection{ROT recovered with the Maxwellian Iteration}\label{sec:2_1}
It is easy to prove that eqs. \eqref{10}$_{1,2}$ can be written as (see \cite{P-R} for more details) 
\begin{align}\label{Belle}
\begin{split}
& V^\alpha_E \, \partial_{\alpha} \, \lambda^E \, + \, T^{\alpha \mu}_E \, \partial_{\alpha} \, \lambda^E_\mu = 0 \, ,  \quad T^{\alpha \beta}_E \, \partial_{\alpha} \, \lambda^E \, + \, A^{\alpha \beta \mu}_E \, \partial_{\alpha} \, \lambda^E_\mu = 0 \, . 
\end{split}
\end{align}
The first one of these equations and the second one contracted with $U_\beta$ give a system whose solution is:
\begin{align}\label{2c}
\begin{split}
& U^\alpha \, \partial_\alpha \lambda^E = - \, \left| \begin{matrix}
\rho && \frac{e}{c^2} \\
&& \\
\frac{e}{c^2} && \rho \, \theta_{0,2}
\end{matrix} \right|^{-1} \, \left| \begin{matrix}
p && \frac{e}{c^2} \\
&& \\
\frac{1}{3} \, \rho \, c^2 \, \theta_{1,2} && \rho \, \theta_{0,2}
\end{matrix} \right| \, h^{\alpha \delta} \, \partial_\alpha \lambda^E_\delta \, , \\
& U^\alpha U^\beta \, \partial_\alpha \lambda^E_\beta = - \, \left| \begin{matrix}
\rho && \frac{e}{c^2} \\
&& \\
\frac{e}{c^2} && \rho \, \theta_{0,2}
\end{matrix} \right|^{-1} \, \left| \begin{matrix}
\rho && p \\
&& \\
\frac{e}{c^2} &&  \frac{1}{3} \, \rho \, c^2 \, \theta_{1,2}
\end{matrix} \right| \, h^{\alpha \delta} \, \partial_\alpha \lambda^E_\delta \, .
\end{split}
\end{align}
It is interesting to note that $ h^{\alpha \mu} \, \partial_{\alpha} \, \lambda^E_\mu = \frac{1}{T} \, h^{\alpha \mu} \, \partial_{\alpha} \, U_\mu= - \, \frac{1}{T}  \, \partial_{\alpha} \, U^\alpha$. The equation \eqref{Belle}$_2$, contracted with $h^\beta_\delta$, allows us to determine 
\begin{align}\label{3c}
h^{\alpha \theta} \, \partial_\alpha \lambda^E = - \, \frac{2}{3} \, \frac{\rho}{p} \, c^2 \, \theta_{1,2} \, h^{\theta ( \alpha } U^{\delta )} \, \partial_\alpha \lambda^E_\delta \, .
\end{align}
Now we consider \eqref{10}$_3$, with use of \eqref{2_b}, \eqref{1}$_2$ and taking into account that \newline 
$q_\gamma= U^\beta \left( T_{\gamma \beta} \, - \, T_{E \, \gamma \beta} \right)$; jointly with $V^\alpha -  V^\alpha_E = 0$, $U_\alpha U_\beta \,  \left( T^{\alpha \beta} -  T^{\alpha \beta}_E \right) = 0$ we obtain the system
\begin{align}\label{15}
\begin{split}
& \sum_{m=0}^{N} U_\alpha \left( A^{\alpha \alpha_1 \cdots \alpha_n \beta_1 \cdots \beta_m}_E \, + \, \frac{3}{c^4 \rho \, \theta_{1,2}} \, g_{\gamma \delta \, U_\beta \,  A^{\alpha \gamma \alpha_1 \cdots \alpha_n}_E A^{\delta \beta \beta_1 \cdots \beta_m}_E} \right) \left( \lambda_{\beta_1 \cdots \beta_m} \, - \, \lambda_{\beta_1 \cdots \beta_m}^E \right) = \\
& = - \, c^2 \tau \, \left( A^{\alpha \alpha_1 \cdots \alpha_n}_E \, \partial_{\alpha} \, \lambda^E \, + \, A^{\alpha \alpha_1 \cdots \alpha_n \mu}_E \, \partial_{\alpha} \, \lambda^E_\mu \right)^{MI} \\
& =  - \,  c^2 \tau \,  \left( A^{\alpha \alpha_1 \cdots \alpha_n}_E \, \partial_{\alpha} \, \lambda^E \, + \, A^{\alpha \alpha_1 \cdots \alpha_n \mu}_E \, \partial_{( \alpha} \, \lambda^E_{\mu )} \right) \, , \, \mbox{for} \quad n=2 \, , \, \cdots \, , N \, , \\
& \sum_{m=0}^{N} A^{\alpha  \beta_1 \cdots \beta_m}_E \left( \lambda_{\beta_1 \cdots \beta_m} \, - \, \lambda_{\beta_1 \cdots \beta_m}^E \right)^{MI} = 0 \, , \\
& \sum_{m=0}^{N} U_\alpha U_\beta \, A^{\alpha \beta \beta_1 \cdots \beta_m}_E  \left( \lambda_{\beta_1 \cdots \beta_m} \, - \, \lambda_{\beta_1 \cdots \beta_m}^E \right)^{MI} = 0 \, .
\end{split}
\end{align}
From this system we can obtain $\left( \lambda_{\beta_1 \cdots \beta_m} \, - \, \lambda_{\beta_1 \cdots \beta_m}^E \right)^{MI} $ as a linear and homogeneous combination of $\partial_{\alpha} \, \lambda^E$, $\partial_{( \alpha} \, \lambda^E_{\mu )}$. By using \eqref{2c} and \eqref{3c}, $\left( \lambda_{\beta_1 \cdots \beta_m} \, - \, \lambda_{\beta_1 \cdots \beta_m}^E \right)^{MI} $ will be a linear and homogeneous combination of \newline 
$h^{\alpha \mu} \, \partial_{\alpha} \, \lambda^E_\mu$, $ h^{\delta \alpha } \, U^{\mu }	\, \partial_{( \alpha} \, \lambda^E_{\mu )}$ and $ h^{\alpha < \delta } \, h^{ \beta >_3 \mu}	\, \partial_{( \alpha} \, \lambda^E_{\mu )}= h^{\alpha \delta } \, h^{ \beta \mu}	\, \partial_{< \alpha} \, \lambda^E_{\mu >_3}$. By substituting these $\left( \lambda_{\beta_1 \cdots \beta_m} \, - \, \lambda_{\beta_1 \cdots \beta_m}^E \right)^{MI} $ in 
\begin{align}\label{15b}
\left( T^{\alpha \beta} -  T^{\alpha \beta}_E \right)^{MI} = - \, \frac{m}{k_B} \sum_{m=0}^{N}  A^{\alpha \beta \beta_1 \cdots \beta_m}_E  \left( \lambda_{\beta_1 \cdots \beta_m} \, - \, \lambda_{\beta_1 \cdots \beta_m}^E \right)^{MI} \, , 
\end{align}
we obtain that $h_{\alpha \beta} \left( T^{\alpha \beta} -  T^{\alpha \beta}_E \right)^{MI}$
is a scalar function which is linear and homogeneous in the independent variables $h^{\alpha \mu} \, \partial_{\alpha} \, \lambda^E_\mu$ (a scalar variable), $ h^{\delta \alpha } \, U^{\mu }	\, \partial_{( \alpha} \, \lambda^E_{\mu )}$ (a 3-dimensional vector variable) and $h^{\alpha \delta } \, h^{ \beta \mu}	\, \partial_{< \alpha} \, \lambda^E_{\mu >_3}$ (a variable which is  a 3-dimensional second order tensor). For the representation theorems (see \cite{Pen-Tro} for more details on the representation theorems) the quantity $h_{\alpha \beta} \left( T^{\alpha \beta} -  T^{\alpha \beta}_E \right)^{MI}$ must be proportional to \newline 
$h^{\alpha \mu} \, \partial_{\alpha} \, \lambda^E_\mu= \frac{- 1}{T} \, \partial_{\alpha} \, U^\alpha$. In this way, \eqref{9}$_1$
is obtained with $\nu_{MI}^N$ instead of $\nu$.
Similarly, $h_{\alpha \delta} \left( T^{\alpha \beta} -  T^{\alpha \beta}_E \right)^{MI}$
is a 3-dimensional vector function which is linear and homogeneous in the same independent variables. So, by the representation theorems, $h_{\alpha \delta} \left( T^{\alpha \beta} -  T^{\alpha \beta}_E \right)^{MI}$ must be proportional to \newline $ 2 \, h^{\delta \alpha } \, U^{\mu }	\, \partial_{( \alpha} \, \lambda^E_{\mu )}= \frac{-c^2}{T^2} \, h^{\delta \alpha } \, \left( \partial_{ \alpha} \, T \, - \, \frac{T}{c^2}  \, U^{\mu }	\, \partial_{\mu} \, U_{\alpha} \right)$.  In this way, \eqref{9}$_2$
is obtained with $\chi_{MI}^N$ instead of $\chi$. \\
Finally, $h_\mu ^{< \beta} h^{\gamma >}_\nu \, \left( T^{\mu \nu} -  T^{\mu \nu}_E \right)^{MI}$
is a 3-dimensional traceless second order tensorial  function which is linear and homogeneous in the same independent variables. By the representation theorems $h_\mu ^{< \beta} h^{\gamma >}_\nu \, \left( T^{\mu \nu} -  T^{\mu \nu}_E \right)^{MI}$ must be proportional to 
$h^\alpha_\beta \, h^\mu_\gamma \, \partial_{< \alpha} \, U_{\mu >} $.
In this way, \eqref{9}$_3$
is obtained with $\mu_{MI}^N$ instead of $\mu$. \\
As examples of this procedure, we will consider the particular cases with  $N=3$ and $N=2$ in the next subsections.

\subsection{The Maxwellian iteration in the case  $N=3$.}\label{sec:2_2}
We begin with the case  $N=3$ because the calculations developed in this case allow us also                                                                                                                                                                                                                                                                                                                                                                                                                                                                                                                                                                                                                                                                                                                                                                                                                                                                                                                                                                                                                                                             to treat the case with  $N=2$ which will be considered in the next subsection. By comparing the results obtained in the case $N=3$ with those obtained in the case $N=2$, it is easy to see that the expressions of the coefficients of bulk viscosity, heat conductivity and shear stress are different. This argument allows us to prove that the expressions of $\nu, \chi$ and $\mu$ obtained with the Maxwellian iteration depend on $N$. So far in literature, this fact has been proved only comparing the case $N=2$ with two of its subsystems. \newline
The equations \eqref{15} and \eqref{15b} for $N=3$ are:
\begin{align}\label{23}
\begin{split}
&U_\alpha \left( A^{\alpha \alpha_1 \cdots \alpha_n}_E \, + \, \frac{3}{c^4 \rho \, \theta_{1,2}} \, g_{\gamma \delta \, U_\beta \,  A^{\alpha \gamma \alpha_1 \cdots \alpha_n}_E A^{\delta \beta}_E} \right) \left( \lambda \, - \, \lambda^E \right)^{MI}  +   \\
& + \, U_\alpha \left( A^{\alpha \alpha_1 \cdots \alpha_n \beta_1}_E \, + \, \frac{3}{c^4 \rho \, \theta_{1,2}} \, g_{\gamma \delta \, U_\beta \,  A^{\alpha \gamma \alpha_1 \cdots \alpha_n}_E A^{\delta \beta \beta_1}_E} \right) \left( \lambda_{\beta_1} \, - \, \lambda_{\beta_1}^E \right)^{MI}  +  \\
& + \, \sum_{m=0}^{N} U_\alpha \left( A^{\alpha \alpha_1 \cdots \alpha_n \beta_1 \beta_2}_E \, + \, \frac{3}{c^4 \rho \, \theta_{1,2}} \, g_{\gamma \delta \, U_\beta \,  A^{\alpha \gamma \alpha_1 \cdots \alpha_n}_E A^{\delta \beta \beta_1 \beta_2}_E} \right) \,  \left( \lambda_{\beta_1 \beta_2}  \right)^{MI} \, +  \\
& + \, U_\alpha \left( A^{\alpha \alpha_1 \cdots \alpha_n \beta_1 \beta_2 \beta_3}_E \, + \, \frac{3}{c^4 \rho \, \theta_{1,2}} \, g_{\gamma \delta \, U_\beta \,  A^{\alpha \gamma \alpha_1 \cdots \alpha_n}_E A^{\delta \beta \beta_1 \beta_2 \beta_3}_E} \right) \, \left( \lambda_{\beta_1 \beta_2 \beta_3} \right)^{MI}  =  \\
& =  -  \, c^2 \tau \,  \left( A^{\alpha \alpha_1 \cdots \alpha_n}_E \, \partial_{\alpha} \, \lambda^E \, + \, A^{\alpha \alpha_1 \cdots \alpha_n \mu}_E \, \partial_{( \alpha} \, \lambda^E_{\mu )} \right) \, , \, \mbox{for} \quad n=2 \, , \, \cdots \, , 3 \, , 
\end{split}
\end{align}
\begin{align*}
 A^{\alpha}_E \left( \lambda \, - \, \lambda^E \right)^{MI}  \, +  \,A^{\alpha  \beta_1}_E \left( \lambda_{\beta_1} \, - \, \lambda_{\beta_1}^E \right)^{MI}  \, + \,  A^{\alpha  \beta_1 \beta_2}_E \left( \lambda_{\beta_1 \beta_2}  \right)^{MI}  \, +  \\
 + \,  A^{\alpha  \beta_1 \beta_2 \beta_3}_E \left( \lambda_{\beta_1 \beta_2 \beta_3} \right)^{MI} = 0 \, , 
\end{align*}
\begin{align*}
U_\alpha U_\beta \, \left[  A^{\alpha \beta}_E \left( \lambda \, - \, \lambda^E \right)^{MI}  \, +  \,A^{\alpha  \beta \beta_1}_E \left( \lambda_{\beta_1} \, - \, \lambda_{\beta_1}^E \right)^{MI}  \, + \,  A^{\alpha  \beta \beta_1 \beta_2}_E \left( \lambda_{\beta_1 \beta_2}  \right)^{MI}  \, +  \right. \\
+ \left. \,  A^{\alpha  \beta \beta_1 \beta_2 \beta_3}_E \left( \lambda_{\beta_1 \beta_2 \beta_3} \right)^{MI} \right] = 0 \, .
\end{align*}
\begin{align*}
\left( T^{\alpha \beta} -  T^{\alpha \beta}_E \right)^{MI} = - \, \frac{m}{k_B} \left[  A^{\alpha \beta}_E \left( \lambda \, - \, \lambda^E \right)^{MI}  \, +  \,A^{\alpha  \beta \beta_1}_E \left( \lambda_{\beta_1} \, - \, \lambda_{\beta_1}^E \right)^{MI}  \, +  \right. \\
+ \left. \,  A^{\alpha  \beta \beta_1 \beta_2}_E \left( \lambda_{\beta_1 \beta_2}  \right)^{MI}  \, + \,  A^{\alpha  \beta \beta_1 \beta_2 \beta_3}_E \left( \lambda_{\beta_1 \beta_2 \beta_3} \right)^{MI} \right] \, . 
\end{align*}
We note that in the case $N=2$ we have to consider \eqref{23}$_1$ only for $n=2$ and to put 
$\left( \lambda_{\beta_1 \beta_2 \beta_3} \right)^{MI} =0 $; so the calculations in this section allow us to consider also the case $N=2$ which is the object of the next section.

{\bf Determination of the bulk viscosity $\nu$.}
Here we consider eqs. \eqref{23}$_1$ with $n=2$ contracted by $\frac{ U_{\alpha_1} U_{\alpha_2}}{\rho \, c^6}$, \eqref{23}$_1$ with $n=3$ contracted by $\frac{ U_{\alpha_1} U_{\alpha_2} U_{\alpha_3}}{\rho \, c^8}$, \eqref{23}$_1$ with $n=2$ contracted by $\frac{ h_{\alpha_1 \alpha_2}}{\rho \, c^4}$, \eqref{23}$_1$ with $n=3$ contracted by $\frac{ h_{\alpha_1 \alpha_2} U_{\alpha_3}}{\rho \, c^6}$, \eqref{23}$_2$  contracted by $\frac{ U_\alpha}{\rho \, c^2}$, \eqref{23}$_3$  divided by $\rho \, c^6$ and \eqref{23}$_4$ contracted by $- \, k_B \, \frac{ h_{\alpha \beta}}{m \, \rho \, c^2}$. \\
So we obtain a system $\sum_{j=1}^{6} a_{ij} X^j = b_i$ constituted by 7 equations in the 6 unknowns $X^1= \left( \lambda \, - \, \lambda^E \right)^{MI}$,  $X^2= U^{\beta_1} \left( \lambda_{\beta_1} \, - \, \lambda_{\beta_1}^E \right)^{MI}$,  $X^3= U^{\beta_1} U^{\beta_2} \left( \lambda_{\beta_1 \beta_2} \right)^{MI}$,  $X^4= U^{\beta_1} U^{\beta_2} U^{\beta_3} \left( \lambda_{\beta_1 \beta_2 \beta_3} \right)^{MI}$, $X^5= c^2 h^{\beta_1 \beta_2} \left( \lambda_{\beta_1 \beta_2} \right)^{MI}$,  $X^6= c^2 h^{\beta_1 \beta_2} U^{\beta_3} \left( \lambda_{\beta_1 \beta_2 \beta_3} \right)^{MI}$ where 
\begin{align}\label{24}
\begin{split}
& a_{1k} = \theta_{0,k+1} \, + \, 3 \, \frac{\theta_{0,3}}{\theta_{1,2}} \, \theta_{0,k} \, , \quad \mbox{for} \quad k=1,2,3,4. \, ; \quad a_{15} = \frac{1}{10} \, \theta_{1,4} \, + \, \frac{1}{2} \, \frac{\theta_{0,3} \, \theta_{1,3} }{\theta_{1,2}} \, ; \\
& a_{16} = \frac{1}{5} \, \theta_{1,5} \, + \, \frac{9}{10} \, \frac{\theta_{0,3} \, \theta_{1,4} }{\theta_{1,2}} \, ; \\ 
& b_1 = - \, \tau \, \left[ \theta_{0,2} \, U^\alpha \partial_{ \alpha}  \, \lambda^E \, + \left( \theta_{0,3} U^\alpha U^\mu \, + \, \frac{1}{6} \, c^2 \, \theta_{1,3} \, h^{\alpha \mu} \right) \partial_{( \alpha} \, \lambda^E_{\mu )} \right] \, ; 
\end{split}
\end{align}
\begin{align*}
\begin{split}
& a_{2k} = \theta_{0,k+2} \, + \, 3 \, \frac{\theta_{0,4}}{\theta_{1,2}} \, \theta_{0,k} \, , \quad \mbox{for} \quad k=1,2,3,4. \, ; \quad a_{25} = \frac{1}{15} \, \theta_{1,5} \, + \, \frac{1}{2} \, \frac{\theta_{0,4} \, \theta_{1,3} }{\theta_{1,2}} \, ; \\
& a_{26} = \frac{1}{7} \, \theta_{1,6} \, + \, \frac{9}{10} \, \frac{\theta_{0,4} \, \theta_{1,4} }{\theta_{1,2}} \, ; \\ 
& b_2 = -  \, \tau \, \left[ \theta_{0,3} \, U^\alpha \partial_{ \alpha}  \, \lambda^E \, + \left( \theta_{0,4} U^\alpha U^\mu \, + \, \frac{3}{10} \, c^2 \, \theta_{1,4} \, h^{\alpha \mu} \right) \partial_{( \alpha} \, \lambda^E_{\mu )} \right] \, ; 
\end{split}
\end{align*}
\begin{align*}
\begin{split}
& a_{31} =  \theta_{1,1} \, + \, \frac{3}{2} \, \frac{\theta_{0,1} \, \theta_{1,3} }{\theta_{1,2}} \, ; \, a_{32} =  \frac{1}{3} \, \theta_{1,3} \, + \, \frac{3}{2} \, \frac{\theta_{0,2} \, \theta_{1,3} }{\theta_{1,2}} \, ; \, a_{33} =  \frac{3}{10} \, \theta_{1,4} \, + \, \frac{3}{2} \, \frac{\theta_{0,3} \, \theta_{1,3} }{\theta_{1,2}} \, ;  \\
& a_{34} =  \frac{4}{15} \, \theta_{1,5} \, + \, \frac{3}{2} \, \frac{\theta_{0,4} \, \theta_{1,3} }{\theta_{1,2}} \, ;  \, a_{35} = \frac{1}{3} \, \theta_{2,4} \, + \, \frac{1}{4} \, \frac{\left( \theta_{1,3} \right)^2 }{\theta_{1,2}} \, ; \\ 
& a_{36} = \frac{1}{3} \, \theta_{2,5} \, + \, \frac{9}{20} \, \frac{\theta_{1,4} \, \theta_{1,3} }{\theta_{1,2}} \, ; \\ 
& b_3 = -  \, \tau \, \left[ \theta_{1,2} \, U^\alpha \partial_{ \alpha}  \, \lambda^E \, + \left( \frac{1}{2} \, \theta_{1,3} U^\alpha U^\mu \, + \, \frac{5}{3} \, c^2 \, \theta_{2,3} \, h^{\alpha \mu} \right) \partial_{( \alpha} \, \lambda^E_{\mu )} \right] \, ; 
\end{split}
\end{align*}
\begin{align*}
\begin{split}
& a_{41} = \frac{1}{2} \,  \theta_{1,3} \, + \, \frac{9}{10} \, \frac{\theta_{0,1} \, \theta_{0,4} }{\theta_{1,2}} \, ; \, a_{42} =  \frac{3}{10} \, \theta_{1,4} \, + \, \frac{9}{10} \, \frac{\theta_{0,2} \, \theta_{0,4} }{\theta_{1,2}} \, ; \, a_{43} =  \frac{1}{5} \, \theta_{1,5} \, + \, \frac{9}{10} \, \frac{\theta_{0,3} \, \theta_{1,4} }{\theta_{1,2}} \, ;  \\
& a_{44} =  \frac{1}{7} \, \theta_{1,6} \, + \, \frac{9}{10} \, \frac{\theta_{0,4} \, \theta_{1,4} }{\theta_{1,2}} \, ;  \, a_{45} = \frac{1}{9} \, \theta_{2,5} \, + \, \frac{3}{20} \, \frac{ \theta_{1,3} \, \theta_{1,4} }{\theta_{1,2}} \, ; \\ 
& a_{46} = \frac{1}{7} \, \theta_{2,6} \, + \, \frac{27}{100} \, \frac{ \left( \theta_{1,4} \right)^2 }{\theta_{1,2}} \, ; \\ 
& b_4 = -  \, \tau \, \left[ \frac{1}{2} \, \theta_{1,3} \, U^\alpha \partial_{ \alpha}  \, \lambda^E \, + \left( \frac{3}{10} \, \theta_{1,4} U^\alpha U^\mu \, + \, \frac{1}{3} \, c^2 \, \theta_{2,4} \, h^{\alpha \mu} \right) \partial_{( \alpha} \, \lambda^E_{\mu )} \right] \, ; 
\end{split}
\end{align*}
\begin{align*}
a_{51} = \theta_{0,0}  \, ; \, a_{52} =  \theta_{0,1} \, ; \, a_{53} =  \theta_{0,2} \, ;  \, a_{54} =  \theta_{0,3} \,  ;  \, a_{55} = \frac{1}{3} \, \theta_{1,2} \, ; \,
 a_{56} = \frac{1}{2} \, \theta_{1,3} \, ; \,  b_5 = 0 \, ; 
\end{align*}
\begin{align*}
a_{61} = \theta_{0,1}  \, ; \, a_{62} =  \theta_{0,2} \, ; \, a_{63} =  \theta_{0,3} \, ;  \, a_{64} =  \theta_{0,4} \,  ;  \, a_{65} = \frac{1}{6} \, \theta_{1,3} \, ; \,
a_{66} = \frac{3}{10} \, \theta_{1,4} \, ; \,  b_6 = 0 \, ; 
\end{align*}
\begin{align*}
\begin{split}
& a_{71} = 3 \, \theta_{1,1}  \, ; \, a_{72} =  \theta_{1,2} \, ; \, a_{73} = \frac{1}{2} \, \theta_{1,3} \, ;  \, a_{74} = \frac{3}{10} \, \theta_{1,4} \,  ;  \, a_{75} = \frac{5}{3} \, \theta_{2,5} \, ; \,
a_{76} = \theta_{2,4} \, ; \\ 
& b_7 =  - \, \frac{ k_B}{m \, \rho \, c^2} \, \left( T^{\alpha \beta} -  T^{\alpha \beta}_E \right)^{MI} \, h_{\alpha \beta} \, ;
\end{split}
\end{align*}
where the expression of $\theta_{k,j}$ has been introduced in \eqref{SC}.
By applying the Rouch\'e-Capelli Theorem, we see that the determinant of the augmented matrix must be zero; so, from this condition, by calling $D_j$ the algebraic complement on the line $i$, column 7 of this matrix, we find 
\begin{align*}
\begin{split}
&  h_{\alpha \beta} \, \left( T^{\alpha \beta} -  T^{\alpha \beta}_E \right)^{MI} = - \, \frac{c^2 m \rho \, \tau }{k_B} \cdot \\
&  \cdot \left\{ \frac{D_1}{D_7} \, \left[ \theta_{0,2} \, U^\alpha \partial_{ \alpha}  \, \lambda^E \, + \left( \theta_{0,3} U^\alpha U^\mu \, + \, \frac{1}{6} \, c^2 \, \theta_{1,3} \, h^{\alpha \mu} \right) \partial_{( \alpha} \, \lambda^E_{\mu )}  \right] \right. + \\
&  + \,  \frac{D_2}{D_7} \, \left[ \theta_{0,3} \, U^\alpha \partial_{ \alpha}  \, \lambda^E \, + \left( \theta_{0,4} U^\alpha U^\mu \, + \, \frac{3}{10} \, c^2 \, \theta_{1,4} \, h^{\alpha \mu} \right) \partial_{( \alpha} \, \lambda^E_{\mu )} \right] + \\
&  + \frac{D_3}{D_7} \, \left[ \theta_{1,2} \, U^\alpha \partial_{ \alpha}  \, \lambda^E \, + \left( \frac{1}{2} \, \theta_{1,3} U^\alpha U^\mu \, + \, \frac{5}{3} \, c^2 \, \theta_{2,3} \, h^{\alpha \mu} \right) \partial_{( \alpha} \, \lambda^E_{\mu )} \right] + \\
&  \left. \, + \,  \frac{D_4}{D_7} \, \left[ \frac{1}{2} \, \theta_{1,3} \, U^\alpha \partial_{ \alpha}  \, \lambda^E \, + \left( \frac{3}{10} \, \theta_{1,4} U^\alpha U^\mu \, + \, \frac{1}{3} \, c^2 \, \theta_{2,4} \, h^{\alpha \mu} \right) \partial_{( \alpha} \, \lambda^E_{\mu )}\right] \right\} \, ,
\end{split}
\end{align*}
i.e., by using \eqref{2c}, 
\begin{align}\label{25}
\begin{split}
&  h_{\alpha \beta} \, \left( T^{\alpha \beta} -  T^{\alpha \beta}_E \right)^{MI} = - \, 3 \, \nu^{MI} \,  \partial_{\alpha} \, U^\alpha \, , \quad \mbox{with} \\
&    \nu^{MI}  =   \frac{c^2 m \rho \, \tau }{3 \, k_B T}  \,  \left[\left(  \frac{D_1}{D_7} \,  \theta_{0,2}   + \,  \frac{D_2}{D_7} \, \theta_{0,3} + \frac{D_3}{D_7} \,  \theta_{1,2
}  \, + \,  \frac{1}{2} \, \theta_{1,3}  \right)  \, \frac{\left| \begin{matrix}
	\frac{p}{\rho} & & \frac{e}{\rho \, c^2} \\
	&& \\
	c^2 \theta_{1,2} & &\theta_{0,2} 
	\end{matrix}  \right|}{\left| \begin{matrix}
	1 & &\frac{e}{\rho \, c^2} \\
	&& \\
	\frac{e}{\rho \, c^2} & & \theta_{0,2} 
	\end{matrix}  \right|} \, \right. +  \\
&  + \, \left( \frac{D_1}{D_7} \,  \frac{1}{6} \, c^2 \, \theta_{1,3} \,  + \,  \frac{D_2}{D_7} \, \frac{3}{10} \, c^2 \, \theta_{1,4} \, + \,
\frac{D_3}{D_7} \,  \frac{5}{3} \, c^2 \, \theta_{2,3} \,  + \,
\frac{D_4}{D_7} \, \frac{1}{3} \, c^2 \, \theta_{2,4} \right)  + \\
& + \left. \, \left( \frac{D_1}{D_7} \,  \theta_{0,3} \, +  \,  \frac{D_2}{D_7} \,  \theta_{0,4}  \, + \,  \frac{D_4}{D_7} \, \frac{3}{10} \, \theta_{1,4} \right) \, \frac{\left| \begin{matrix}
	1 & & \frac{p}{\rho} \\
	&& \\
	\frac{e}{\rho \, c^2} & & c^2 \, \theta_{1,2} 
	\end{matrix}  \right|}{\left| \begin{matrix}
	1 & &\frac{e}{\rho \, c^2} \\
	&& \\
	\frac{e}{\rho \, c^2} & & \theta_{0,2} 
	\end{matrix}  \right|} \right] \, .
\end{split}
\end{align}

{\bf Determination of the heat conductivity $\chi$.}
We consider now eqs. \eqref{23}$_1$ with $n=2$ contracted by $\frac{ h^\theta_{\alpha_1} U_{\alpha_2}}{\rho \, c^6}$, \eqref{23}$_1$ with $n=3$ contracted by $\frac{ h^\theta_{\alpha_1} U_{\alpha_2} U_{\alpha_3}}{\rho \, c^8}$,  \eqref{23}$_1$ with $n=3$ contracted by $\frac{ h_{\alpha_1 \alpha_2} h^\theta_{\alpha_3}}{\rho \, c^6}$, \eqref{23}$_2$  contracted by $\frac{ h^\theta_\alpha}{- \, \rho \, c^2}$ and \eqref{23}$_4$ contracted by $ k_B \, \frac{ h^\theta_\alpha U\beta}{\rho \, c^4}$. \\
So we obtain a system $\sum_{j=1}^{4} b_{ij} X^{j \theta} = b_{i}^\theta$ constituted by 5 equations in the 4 unknowns $X^{1 \theta}=  h^{\theta \beta_1} \left( \lambda_{\beta_1} \, - \, \lambda_{\beta_1}^E \right)^{MI}$,  $X^{2 \theta}= h^{\theta \beta_1} U^{\beta_2} \left( \lambda_{\beta_1 \beta_2} \right)^{MI}$,  $X^{3 \theta}= h^{\theta \beta_1} U^{\beta_2} U^{\beta_3} \left( \lambda_{\beta_1 \beta_2 \beta_3} \right)^{MI}$, $X^{4 \theta}= c^2 h^{\theta \beta_1}  h^{\beta_2 \beta_3} \, \left( \lambda_{\beta_1 \beta_2 \beta_3} \right)^{MI}$ with coefficients
\begin{align}\label{26}
\begin{split}
& b_{11} = - \, \frac{1}{3} \, \theta_{1,3}  \, ;  \quad b_{12} = - \, \frac{1}{5} \, \theta_{1,4} \, + \, \frac{1}{6} \, \frac{\left( \theta_{1,3} \right)^2 }{\theta_{1,2}} \, ; \\ 
& b_{13} =  - \, \frac{1}{5} \, \theta_{1,5} \, + \, \frac{3}{20} \, \frac{\theta_{1,3} \, \theta_{1,4} }{\theta_{1,2}} \, ; \quad b_{14} = - \, \frac{1}{15} \, \theta_{2,5} \, + \, \frac{1}{10} \, \frac{ \theta_{1,3} \, \theta_{2,4} }{\theta_{1,2}} \, ;  \\ 
& b_{1}^\theta =  \tau \, \left( \frac{1}{3} \, \theta_{1,2} \, h^{\theta \alpha} \partial_{ \alpha}  \, \lambda^E \, + \, \frac{1}{3}  \, \theta_{1,3} \, h^{\theta \alpha} U^\mu \, \partial_{( \alpha} \, \lambda^E_{\mu )} \right) \, ; 
\end{split}
\end{align}
\begin{align*}
\begin{split}
& b_{21} = - \, \frac{1}{5} \, \theta_{1,4}  \, ;  \quad b_{22} = - \, \frac{2}{15} \, \theta_{1,5} \, + \, \frac{1}{10} \, \frac{ \theta_{1,4} \, \theta_{1,3}  }{\theta_{1,2}} \, ; \\ 
& b_{23} =  - \, \frac{1}{7} \, \theta_{1,6} \, + \, \frac{9}{100} \, \frac{ \left( \theta_{1,4} \right)^2 }{\theta_{1,2}} \, ; \quad b_{24} = - \, \frac{1}{35} \, \theta_{2,6} \, + \, \frac{3}{50} \, \frac{ \theta_{1,4} \, \theta_{2,4} }{\theta_{1,2}} \, ;  \\ 
& b_{2}^\theta =  \tau \, \left( \frac{1}{6} \, \theta_{1,3} \, h^{\theta \alpha} \partial_{ \alpha}  \, \lambda^E \, + \, \frac{1}{5}  \, \theta_{1,4} \, h^{\theta \alpha} U^\mu \, \partial_{( \alpha} \, \lambda^E_{\mu )} \right) \, ; 
\end{split}
\end{align*} 
\begin{align*}
\begin{split}
& b_{31} = - \, \frac{2}{3} \, \theta_{2,4}  \, ;  \quad b_{32} = - \, \frac{4}{15} \, \theta_{2,5} \, + \, \frac{1}{3} \, \frac{ \theta_{2,4} \, \theta_{1,3}  }{\theta_{1,2}} \, ; \\ 
& b_{33} =  - \, \frac{1}{7} \, \theta_{2,6} \, + \, \frac{3}{10} \, \frac{  \theta_{1,4} \, \theta_{2,4}  }{\theta_{1,2}} \, ; \quad b_{34} = - \, \frac{1}{5} \, \theta_{3,6} \, + \, \frac{1}{5} \, \frac{ \left( \theta_{2,4} \right)^2 }{\theta_{1,2}} \, ;  \\ 
& b_{3}^\theta = \, \tau \, \left(  \theta_{2,3} \, h^{\theta \alpha} \partial_{ \alpha}  \, \lambda^E \, + \, \frac{2}{3}  \, \theta_{2,4} \, h^{\theta \alpha} U^\mu \, \partial_{( \alpha} \, \lambda^E_{\mu )} \right) \, ; 
\end{split}
\end{align*}
\begin{align*}
 b_{41} =  \theta_{1,1}  \, ;  \quad b_{42} =  \frac{2}{3} \, \theta_{1,2}  \, ; \quad 
 b_{43} =   \frac{1}{2} \, \theta_{1,3} \,  ; \quad b_{44} =  \theta_{2,3} \,  ;  \quad 
 b_{4}^\theta =0 \, ; 
\end{align*}
\begin{align*}
\begin{split}
& b_{51} =  \frac{1}{3} \, \theta_{1,2}  \, ;  \quad b_{52} =  \frac{1}{3} \, \theta_{1,3}  \, ; \quad b_{53} =  \frac{3}{10} \, \theta_{1,4}  \, ; \quad b_{54} =  \frac{1}{5} \, \theta_{2,4} \, ;  \\ 
& b_{5}^\theta =\frac{k_B}{m \, \rho \, c^4} \, h^\theta_\alpha U_\beta \,  \left( T^{\alpha \beta} \, - \, T^{\alpha \beta}_E \right)^{MI} \, ; 
\end{split}
\end{align*}
By applying the Rouch\'e-Capelli Theorem, we see that the determinant of the augmented matrix must be zero; so, from this condition by calling $M_j$ the algebraic complement on the line $i$, column 5 of this matrix, we find 
\begin{align*}
\begin{split}
&   h^\theta_\alpha U_\beta \,  \left( T^{\alpha \beta} \, - \, T^{\alpha \beta}_E \right)^{MI} = - \, \frac{c^4 m \rho \, \tau}{k_B} \cdot \\
&  \cdot \left[ \frac{M_1}{M_5} \,   \left( \frac{1}{3} \, \theta_{1,2} \, h^{\theta \alpha} \partial_{ \alpha}  \, \lambda^E \, + \, \frac{1}{3}  \, \theta_{1,3} \, h^{\theta \alpha} U^\mu \, \partial_{( \alpha} \, \lambda^E_{\mu )} \right)    \right. + \\
&  + \,  \frac{M_2}{M_5} \,  \left( \frac{1}{6} \, \theta_{1,3} \, h^{\theta \alpha} \partial_{ \alpha}  \, \lambda^E \, + \, \frac{1}{5}  \, \theta_{1,4} \, h^{\theta \alpha} U^\mu \, \partial_{( \alpha} \, \lambda^E_{\mu )} \right) + \\
&  + \frac{M_3}{M_5} \, \left.  \left(  \theta_{2,3} \, h^{\theta \alpha} \partial_{ \alpha}  \, \lambda^E \, + \, \frac{2}{3}  \, \theta_{2,4} \, h^{\theta \alpha} U^\mu \, \partial_{( \alpha} \, \lambda^E_{\mu )} \right) \right]  \, ,
\end{split}
\end{align*}
i.e., by using \eqref{2c}, 
\begin{align}\label{27}
\begin{split}
   q^\theta &=  - \, \chi^{MI} \, h^{\alpha \theta}
\left( \partial_{\alpha} \, T \, - \, \frac{T}{c^2} \, U^\mu \, \partial_\mu \, U_\alpha \right) \, , \mbox{with} \\ 
  \chi^{MI} &=  \frac{c^6 m \rho \, \tau}{2 \, k_B T^2} \left[ \frac{M_1}{M_5} \,   \left( - \, \frac{2}{3} \, \frac{\rho \, c^2}{p} \, \left( \theta_{1,2} \right)^2 \, + \, \frac{1}{3}  \, \theta_{1,3} \right)    \right. + \\
&  + \,  \frac{M_2}{M_5} \,  \left( - \, \frac{1}{3} \,  \frac{\rho \, c^2}{p} \, \theta_{1,2} \, \theta_{1,3}  \, + \, \frac{1}{5}  \, \theta_{1,4}  \right) + \, \frac{M_3}{M_5} \, \left.  \left(  - \, 2 \,  \frac{\rho \, c^2}{p} \, \theta_{1,2} \, \theta_{2,3}  \, + \, \frac{2}{3}  \, \theta_{2,4} \right) \right]   \, ,
\end{split}
\end{align}

{\bf Determination of the shear viscosity $\mu$.}
Let us consider now eqs. \eqref{23}$_1$ with $n=2$ contracted by $\frac{ h_{\alpha_1 < \theta} h_{\psi > \alpha_2}}{\rho \, c^6}$, \eqref{23}$_1$ with $n=3$ contracted by $\frac{ h_{\alpha_1 < \theta} h_{\psi > \alpha_2} U_{\alpha_3}}{\rho \, c^8}$  and \eqref{23}$_4$ contracted by $\frac{ h_{\alpha < \theta} h_{\psi > \beta}}{\rho \, c^4}$. \\
So we obtain a system $\sum_{j=1}^{2} c_{ij} X^j_{< \theta \psi>} = b_{i < \theta \psi>}$ constituted by 3 equations in the 2 unknowns  $X^1_{< \theta \psi >}= h^{\beta_1}_{< \theta } h^{\beta_2}_{\psi >} \left( \lambda_{\beta_1 \beta_2} \right)^{MI}$, $X^2_{< \theta \psi >}= h^{\beta_1}_{< \theta } h^{\beta_2}_{\psi >} U^{\beta_3} \left( \lambda_{\beta_1 \beta_2 \beta_3} \right)^{MI}$,  with coefficients
\begin{align}\label{28}
 c_{11} =  \frac{2}{15} \, \theta_{2,4}  \, ;  \quad c_{12} =  \frac{2}{15} \, \theta_{2,5}  \, ; \quad b_{1 < \theta \psi>} = -  \, \frac{2}{3}  \, \tau \, \theta_{2,3} \,  \partial_{< \theta} \, \lambda^E_{\psi >_3}  \, ; 
\end{align}
\begin{align*}
c_{21} =  \frac{2}{45} \, \theta_{2,5}  \, ;  \quad c_{22} =  \frac{1}{35} \, \theta_{2,6}  \, ; \quad b_{2 < \theta \psi>} = -  \, \frac{2}{15} \, \tau \, \theta_{2,4} \,  \partial_{< \theta} \, \lambda^E_{\psi >_3}  \, ; 
\end{align*}
\begin{align*}
c_{31} =  \frac{2}{3} \, \theta_{2,3}  \, ;  \quad c_{32} =  \frac{2}{5} \, \theta_{2,4}  \, ; \quad b_{3 < \theta \psi>} = - \, \frac{k_B}{m \, c^4 \rho}  \, h_{\alpha < \theta} h_{\psi > \beta} \left( T^{\alpha \beta} \, - \, T^{\alpha \beta}_E \right)^{MI} \, . 
\end{align*}
From this system we obtain 
\begin{align}\label{29}
\begin{split}
&   t_{ < \theta \psi>} =   2 \, \mu^{MI} \,  \partial_{< \theta} \, U_{\psi >} \, , \mbox{with} \\ 
&  \mu^{MI} =  - \, \frac{c^4 m \, \rho \, \tau}{2 \, k_B T} \, \frac{\left| \begin{matrix}
	\frac{2}{15} \, \theta_{2,4}  &&  \frac{2}{15} \, \theta_{2,5}  && -  \, \frac{2}{3} \,   \theta_{2,3} \\
	&&&& \\
\frac{2}{45} \, \theta_{2,5}  &&  \frac{1}{35} \, \theta_{2,6}  && -  \, \frac{2}{15} \,   \theta_{2,4} \\
	&&&& \\
\frac{2}{3} \, \theta_{2,3}  &&  \frac{2}{5} \, \theta_{2,4}  && 0
	\end{matrix}\right|}{\left| \begin{matrix}
	\frac{2}{15} \, \theta_{2,4}  &&  \frac{2}{15} \, \theta_{2,5}   \\
	&& \\
	\frac{2}{45} \, \theta_{2,5}  &&  \frac{1}{35} \, \theta_{2,6}  
	\end{matrix}\right|}  \, .
\end{split}
\end{align}
Finally, to complete  our analysis, we have also to consider eq.  \eqref{23}$_1$ with $n=3$ contracted by $h_{\alpha_1}^{ < \theta} h^\psi_{\alpha_2} h_{\alpha_3}^{ \theta >}$. But this step will give only $ \lambda_{< \beta_1 \beta_2 \beta_3 >_3}$ which does not appear in $\left( T^{\alpha \beta} \, - \, T^{\alpha \beta}_E \right)^{MI}$.

\subsection{The Maxwellian iteration in the case  $N=2$.}\label{sec:2_3}
For this case  we have to consider \eqref{23}$_1$ only for $n=2$ and put 
$\left( \lambda_{\beta_1 \beta_2 \beta_3} \right)^{MI} =0 $ in all the equations. 
For example, let us see what happens for the determination of the bulk viscosity.

{\bf Determination of the bulk viscosity $\nu$.}
We consider here eqs. \eqref{23}$_1$ with $n=2$ contracted by $\frac{ U_{\alpha_1} U_{\alpha_2}}{\rho \, c^6}$,  \eqref{23}$_1$ with $n=2$ contracted by $\frac{ h_{\alpha_1 \alpha_2}}{\rho \, c^4}$,  \eqref{23}$_2$  contracted by $\frac{ U_\alpha}{\rho \, c^2}$, \eqref{23}$_3$  divided by $\rho \, c^6$ and \eqref{23}$_4$ contracted by $- \, k_B \, \frac{ h_{\alpha \beta}}{m \, \rho \, c^2}$. \\
So we obtain a system composed by 5 equations in the 6 unknowns $X^1= \left( \lambda \, - \, \lambda^E \right)^{MI}$,  $X^2= U^{\beta_1} \left( \lambda_{\beta_1} \, - \, \lambda_{\beta_1}^E \right)^{MI}$,  $X^3= U^{\beta_1} U^{\beta_2} \left( \lambda_{\beta_1 \beta_2} \right)^{MI}$,   $X^4= c^2 h^{\beta_1 \beta_2} \left( \lambda_{\beta_1 \beta_2} \right)^{MI}$. The augmented matrix can be obtained by cutting the rows 2 and 4 and columns 4 and 6 from the augmented matrix introduced in subsection 4.1 and its determinant is given by:
\begin{align*}
\left| \begin{matrix}
a_{11} \quad & a_{12} \quad & a_{13}  \quad &  a_{15} \quad  & b_1 \\
a_{31} \quad & a_{32} \quad & a_{33}  \quad &  a_{35}  \quad & b_3 \\
a_{51} \quad & a_{52} \quad & a_{53}  \quad &  a_{55} \quad & b_5 \\
a_{61} \quad & a_{62} \quad & a_{63}  \quad &  a_{65}  \quad & b_6 \\
a_{71} \quad & a_{72} \quad & a_{73}  \quad &  a_{75}  \quad & b_7 
\end{matrix} \right| = 0 \, , 
\end{align*}
where the expressions of $a_{ij}$ and $b_i$ are the same of the case $N=3$.\newline By calling $D_j$ the algebraic complements of the row $i$, column 5 of the preceding matrix, we find 
\begin{align*}
\begin{split}
& \left( T^{\alpha \beta} -  T^{\alpha \beta}_E \right)^{MI} \, h_{\alpha \beta}   = - \, \frac{m \, \rho \, c^2 \, \tau }{ k_B} \,\left( \frac{ \theta_{0,2} \, D_1 \, + \, \theta_{1,2} \, D_2 }{D_5} \, \, U^\alpha \partial_{ \alpha}  \, \lambda^E + \right. \\
&   + \, \frac{   \theta_{0,3}   \, D_1 \, + \,   \frac{1}{2} \, \theta_{1,3}    \, D_2 }{D_5} \, U^\alpha U^\mu \, \partial_{( \alpha} \, \lambda^E_{\mu )} \, + \, c^2 \, \left. \frac{ \frac{1}{6} \, \theta_{1,3}  \, D_1 \, +  \,  \frac{5}{3} \,  \theta_{2,3} \, D_2 }{D_5} \, \, h^{\alpha \mu}  \partial_{( \alpha} \, \lambda^E_{\mu )} \right) \, ,
\end{split}
\end{align*}
i.e., by using \eqref{2c},  
\begin{align}\label{30}
\begin{split}
& \left( T^{\alpha \beta} -  T^{\alpha \beta}_E \right)^{MI} \, h_{\alpha \beta}   =  - \, 3 \, \nu^{2MI} \,  \partial_{\alpha} \, U^\alpha  \, , \quad \mbox{where}  \\
& \nu^{2MI} =   \frac{c^2 m \, \rho \, \tau }{3 \, k_B T}   \,  \left( \frac{ \theta_{0,2} \, D_1 \, + \, \theta_{1,2} \, D_2 }{D_5} \,  \frac{\left| \begin{matrix}
	\frac{p}{\rho} & & \frac{e}{\rho \, c^2} \\
	&& \\
	c^2 \theta_{1,2} & &\theta_{0,2} 
	\end{matrix}  \right|}{\left| \begin{matrix}
	1 & &\frac{e}{\rho \, c^2} \\
	&& \\
	\frac{e}{\rho \, c^2} & & \theta_{0,2} 
	\end{matrix}  \right|}  \, + \right. \\
&  + \, c^2 \, \left. \frac{ \frac{1}{6} \, \theta_{1,3}  \, D_1 \, +  \,  \frac{5}{3} \,  \theta_{2,3} \, D_2 }{D_5} + \, \frac{\theta_{0,3}   \, D_1 \, + \,   \frac{1}{2} \, \theta_{1,3}    \, D_2 }{D_5} \, \frac{\left| \begin{matrix}
	1 & & \frac{p}{\rho} \\
	&& \\
	\frac{e}{\rho \, c^2} & & c^2 \, \theta_{1,2} 
	\end{matrix}  \right|}{\left| \begin{matrix}
	1 & &\frac{e}{\rho \, c^2} \\
	&& \\
	\frac{e}{\rho \, c^2} & & \theta_{0,2} 
	\end{matrix}  \right|} \,   \, \, \right) \, .
\end{split}
\end{align}
It is evident that this value of $\nu$ is different from that found in \eqref{25} when $N=3$.

{\bf Determination of the heat conductivity $\chi$.}
We consider now eqs. \eqref{23}$_1$ with $n=2$ contracted by $\frac{ h^\theta_{\alpha_1} U_{\alpha_2}}{\rho \, c^6}$,   \eqref{23}$_2$  contracted by $\frac{ h^\theta_\alpha}{- \, \rho \, c^2}$ and \eqref{23}$_4$ contracted by $ k_B \, \frac{ h^\theta_\alpha U\beta}{\rho \, c^4}$.
We obtain a system $\sum_{j=1}^{2} b_{ij} X^{j \theta} = b_{i}^\theta$ constituted by 3 equations in the 2 unknowns $X^{1 \theta}=  h^{\theta \beta_1} \left( \lambda_{\beta_1} \, - \, \lambda_{\beta_1}^E \right)^{MI}$ and $X^{2 \theta}= h^{\theta \beta_1} U^{\beta_2} \left( \lambda_{\beta_1 \beta_2} \right)^{MI}$. The augmented matrix can be obtained by eliminating the rows 2 and 3 and the columns 3 and 4 from the augmented matrix introduced in subsection 4.2, so its determinant is given by:
\begin{align*}
\left| \begin{matrix}
b_{11} \quad & b_{12} \quad   & b_1^\theta \\
b_{41} \quad & b_{42} \quad   & b_4^\theta \\
b_{51} \quad & b_{52} \quad   & b_5^\theta 
\end{matrix} \right| = 0 \, , 
\end{align*}
where $b_{ij}$ and $b_i^\theta$ are the same of the case $N=3$.  By calling $D_j$ the algebraic complements of the line $i$, column 3 of the preceding matrix,  we find 
\begin{align*}
\frac{k_B}{m \, \rho \, c^4} \, h^\theta_\alpha U_\beta \,  \left( T^{\alpha \beta} \, - \, T^{\alpha \beta}_E \right)^{MI}   =  - \, \frac{  \tau \, \left( \frac{1}{3} \, \theta_{1,2} \, h^{\theta \alpha} \partial_{ \alpha}  \, \lambda^E \, + \, \frac{1}{3}  \, \theta_{1,3} \, h^{\theta \alpha} U^\mu \, \partial_{( \alpha} \, \lambda^E_{\mu )} \right) \, D_1 }{D_3} \, , 
\end{align*}
i.e., by using \eqref{2c}, 
\begin{align}\label{31}
\begin{split}
&   q^\theta =  - \, \chi^{2MI} \, h^{\alpha \theta}
\left( \partial_{\alpha} \, T \, - \, \frac{T}{c^2} \, U^\mu \, \partial_\mu \, U_\alpha \right) \, , \mbox{where} \\ 
&  \chi^{2MI} = \frac{m \, \rho \, c^6 \tau}{2 \, k_B T^2} \, \frac{D_1}{D_3} \, \left( - \, \frac{2}{3} \, \rho \, c^2 \, \frac{\left( \theta_{1,2} \right)^2 }{p}   \, + \, 
\frac{1}{3}  \, \theta_{1,3} \right) \, ,
\end{split}
\end{align}
It is evident that this value of $\chi$ is different from the value obtained by using the expression of $\chi$ found in \eqref{27} in the case $N=3$.

{\bf Determination of the shear viscosity $\mu$.}
Let us consider now eqs. \eqref{23}$_1$ with $n=2$ contracted by $\frac{ h_{\alpha_1 < \theta} h_{\psi > \alpha_2}}{\rho \, c^6}$  and \eqref{23}$_4$ contracted by $\frac{ h_{\alpha < \theta} h_{\psi > \beta}}{\rho \, c^4}$.
We obtain a system $\sum_{j=1}^{1} c_{ij} X^j_{< \theta \psi>} = b_{i < \theta \psi>}$ constituted by 2 equations in the 1 unknown  $X^1_{< \theta \psi >}= h^{\beta_1}_{< \theta } h^{\beta_2}_{\psi >} \left( \lambda_{\beta_1 \beta_2} \right)^{MI}$. 
The augmented matrix can be obtained by eliminating row 2 and column 2 from the augmented matrix introduced in subsection 4.3 and its determinant is given by:
\begin{align*}
\left| \begin{matrix}
c_{11} \quad   & b_{1 < \theta \psi>} \\
c_{31} \quad & b_{3 < \theta \psi>}
\end{matrix} \right| = 0 \, , 
\end{align*}
where $c_{ij}$ and $b_{i < \theta \psi>}$ are the same of the case $N=3$.  From this equation we find 
\begin{align}\label{number_sh}
 t_{ < \theta \psi>}  =  2 \, \mu^{MI} \,  \partial_{< \theta} \, U_{\psi >} \, , \quad \mbox{with} \\
 \mu^{MI} \,  = \frac{1}{3} \, \frac{m \, c^4 \rho \, \tau}{k_B T}  \, \frac{c_{31}}{c_{11}}  \, \theta_{2,3} = \frac{5}{3} \, \frac{m \, c^4 \rho \, \tau}{k_B T}  \, \frac{\theta_{2,3} }{\theta_{2,4}  }  \, \theta_{2,3}
\, , 
\end{align}
It is evident that this value of $\mu$ is different from the value of $\mu$ furnished by \eqref{29} when $N=3$.
\newline
These transport coefficients are the same ones of \cite{CPTR1} if we take into account that in \cite{CPTR1} the authors call $\omega= \frac{e}{\rho \, c^2}$  (see also eq. $(12)_2$) and the quantities $B_q$, $B_2^\pi$, $B^t$ present in \cite{CPTR1} whose expressions are reported 
in equation (44) in terms also of $C_5$ which is described in eq. (34)$_2$, the matrices $N^\pi$ and $D_4$ which are given in the equation before equation (30) and the matrices $N_3$ and $D_3$ which are given  in the equation after (32)).

\section{The Chapman-Enskog Method}\label{sec3}
This method can be found in the articles \cite{Ensk}, \cite{Chap}  and has been further explained in \cite{CK}.
We describe how this method works by enclosing the full expression of the production term which was found in \cite{PR} and modified in \cite{CPTR1}. In particular, the method starts by considering the following  equations
\begin{align}\label{11}
p^\alpha \, \partial_\alpha \, f = Q \, , \quad \partial_\alpha V^{\alpha} = 0 \, , \quad
\partial_\alpha T^{\alpha \beta} = 0 \, , 
\end{align}
i.e., the Boltzmann equation and the conservation laws of mass and momentum-energy.

Then the following steps have to be followed:
\begin{itemize}
\item[1] The eqs. \eqref{11} are considered, but with their left hand sides calculated at equilibrium and their right hand sides at first order with respect to equilibrium, i.e. 
\begin{align}\label{12}
\begin{split}
& p^\alpha \, \partial_\alpha \, f_E = Q^{(OT)}= \frac{U^\mu p_\mu}{c^2 \tau}\left[\left(f_E-f\right)^{OT}-f_E \, p^\gamma q_\gamma^{OT} \frac{3}{m \, c^4 \rho \, \theta_{1,2}} \, \left(1+\frac{\mathcal{I}}{m c^2}\right) \right] \, . \\
& \\
& \partial_\alpha V^{\alpha}_E = 0 \, , \quad
\partial_\alpha T^{\alpha \beta}_E = 0 \, , 
\end{split}
\end{align}
where the superscript $OT$ denotes that these quantities are the first iterates defined with this approach. 
\item[2] The deviation of the distribution function from its value at equilibrium is calculated in terms of $\partial_{\alpha} \lambda^E$ and $\partial_{\alpha} \lambda_\mu^E$ from \eqref{12}$_1$  and used in eqs. \eqref{2_b}$_{1,2}$ with $n=0$ and $n=1$. Obviously, in this way $I=0$, $I^{\alpha_1}=0$ are obtained, thus respecting the conservation laws of mass and of momentum-energy.
\item[3] The quantities $\partial_{\alpha} \lambda^E$ and $U^\alpha U^\mu \partial_{\alpha} \lambda_\mu^E$ are calculated from \eqref{12}$_{2,3}$ and substituted in the expression of $A^{\alpha} \, - \, A^{\alpha}_E$, $T^{\alpha \beta} \, - \, T^{\alpha \beta}_E$ obtained in the previous step. 
\end{itemize}
 
 We note that in the expression of $Q$ in \cite{CK} there are 6 unknown scalars $a_i$ with $i=0, \cdots 5$ which have to be determined.  From the third line on page 116 of \cite{CK}, by imposing 
$V^\alpha -  V^\alpha_E = 0$, $ e \, - \, e_E = \frac{U_\alpha U_\beta}{c^2} \left( T^{\alpha \beta} -  T^{\alpha \beta}_E \right) = 0$ the authors find $a_0$, $a_1$, $a_3$. By imposing 
$V^\alpha -  V^\alpha_E = 0$, $ e \, - \, e_E = \frac{U_\alpha U_\beta}{c^2} \left( T^{\alpha \beta} -  T^{\alpha \beta}_E \right) = 0$ we find simply that $q_\gamma$ is constrained by $q_\gamma \, U^\gamma =0$. So, $q_\gamma$ replaces the remaining part of Cercignani-Kremer's unknown scalars $a_2$, $a_4$, $a_5$. It is interesting that in 
a model with 14 or more moments, $q_\gamma$ becomes exactly the heat flux density. 
In a model with six moments, there is no heat flux; in this case $q_\gamma$ remains a mathematical tool as the scalars $a_2$, $a_4$, $a_5$ of the Cercignani-Kremer method. But it cannot be eliminated, otherwise the zero deviation of $V^\alpha$ from its value at equilibrium  would be lost. 

%This is the mistake in eq. (45) of \cite{OMS}, altough it does not affect the other results of this article. 

\subsection{ROT recovered with the Chapman-Enskog Method}\label{sec3_1}
In this subsection we apply the Chapman-Enskog method to the equations of polyatomic gases with an arbitrary number $N$.  In this way we will find \eqref{9} of ROT, with particular expressions $\nu^{OT}$,  
$\chi^{OT}$, $\mu^{OT}$ of the bulk viscosity $\nu$, the heat conductivity $\chi$ and the shear viscosity $\mu$ and we will show that all these coefficients do not depend on $N$.\newline

We have to consider the equations 
\begin{align}\label{1c}
\begin{split}
& f \, - \, f_E = \frac{c^2 \tau}{k_B U^\mu p_\mu} \, f_E \, p^\delta 
\left[ m  \, \partial_\delta \, \lambda^E \, +  \left( 1+\frac{\mathcal{I}}{m c^2} \right) \, p^\nu \, \partial_\delta \, \lambda^E_\nu \right]  - \, 3 \, f_E \, p^\mu q_\mu \frac{1+\frac{\mathcal{I}}{m c^2}}{m \, c^4 \rho \, \theta_{1,2}} \, , \\
& V^\alpha \, - \, V^\alpha _E =0 \, , \, U_\alpha U_\beta \left( T^{\alpha \beta} \, - \, T^{\alpha \beta} _E \right) =0 \, , \, \partial_\alpha V^\alpha_E = 0 \, , \, \partial_\alpha T^{\alpha \beta} _E  =0  \, .
\end{split}
\end{align}
The equations \eqref{1c}$_4$ and \eqref{1c}$_5$ are exactly the equations \eqref{Belle} of the MI approach and so the solution of these equations is given by \eqref{2c} and \eqref{3c}. 

Let us now consider equation \eqref{1c}$_2$ contracted with $\frac{U_\alpha}{c^2}$. By using \eqref{1c}$_1$ contracted with $m \, c \, \varphi (\mathcal{I}) \, \frac{U_\alpha}{c^2} \, p^\alpha$ and integrated in $d \, \mathcal{I} \, d \, \vec{P}$ it becomes 
\begin{align*}
0= \frac{U_\alpha}{c^2} \left( V^\alpha \, - \, V^\alpha _E \right)  = \frac{m \, \tau}{k_B} \left( V^\alpha_E \, \partial_\alpha \lambda^E \, + \, T^{\alpha \delta} _E \, \partial_\alpha \lambda^E_\delta \right) \, - \, \frac{3}{\rho \, c^6 \, \theta_{1,2}} \,  U_\alpha \, q_\mu \, T^{\alpha \mu}_E \, ,
\end{align*}
which is an identity for eqs. \eqref{2c} (see also the first equation after \eqref{1c}). \\
To impose  eq. \eqref{1c}$_2$ contracted with $h_\alpha^\theta$, we need the tensors 
\eqref{A0} and their representations \eqref{A1} of the appendix. By using \eqref{1c}$_1$ contracted with $m \, c \, \varphi (\mathcal{I}) \, h_\alpha^\theta \, p^\alpha$ and integrated in $d \, \mathcal{I} \, d \, \vec{P}$ we find 
\begin{align*}
0= h_\alpha^\theta \left( V^\alpha \, - \, V^\alpha _E \right)  = \frac{m \, \tau}{k_B} \left( h_\alpha^\theta \, A^{* \alpha \delta} \, \partial_\delta \lambda^E \, + \, h_\alpha^\theta \, A^{* \alpha \delta \nu} \, \partial_\delta \lambda^E_\nu \right) \, - \, \frac{3}{\rho \, c^4 \, \theta_{1,2}} \,  h_\alpha^\theta \, q_\mu \, T^{\alpha \mu}_E \, ,
\end{align*}
from which we desume
\begin{align}\label{4c}
& q^{\theta} = - \, \frac{m \, \tau c^6}{3 \, k_B} \, \frac{\rho^2}{p} \, \theta_{1,2} \left( \theta_{1,1}^* \, h^{\alpha \theta} \, \partial_\alpha \lambda^E + \, \frac{2}{3} \, \theta_{1,2}^* \, h^{\theta ( \delta} U^{\nu )} \, \partial_\delta \lambda^E_\nu \right) = - \, \chi \, h^{\theta \alpha} \, \left( \partial_{\alpha} \, T \, - \, \frac{T}{c^2} \, U^\mu \, \partial_\mu \, U_\alpha \right) \, , \\
& \mbox{with} \quad \chi = - \, \frac{m \, \tau c^8}{9 \, k_B T^2} \, \frac{\rho^2}{p} \, \theta_{1,2} \left( \theta_{1,2}^* \,  - \, \frac{\rho \, c^2}{p} \, \theta_{1,2} \theta_{1,1}^* \right) \,  , \nonumber
\end{align}
where in the last passage we have used \eqref{3c} and $\lambda_\nu^E = \frac{U\nu}{T}$ . We see here that  $q^\theta$, replaces the  Cercignani-Kremer's scalars which did not have a clear physical meaning. They cannot simply be put equal to zero (as in \cite{OMS}), otherwise the physical requirement $V^\alpha - V^\alpha_E=0$ would be violated.  \\
We now impose eq. \eqref{1c}$_3$, by using \eqref{1c}$_1$ contracted with $U_\alpha U_\beta \, c \, p^\alpha p^\beta \left( 1+\frac{\mathcal{I}}{m c^2} \right) \, \varphi (\mathcal{I})$ and integrated in $d \, \mathcal{I} \, d \, \vec{P}$; we find 
\begin{align*}
0= \frac{m \, c^2 \tau}{k_B} \left( U_\beta T^{\beta \delta}_E   \, \partial_\delta \lambda^E +  U_\beta A^{\beta \delta \nu}_E   \, \partial_\delta \lambda^E_\nu \right) - \, \frac{3}{c^4 \rho \, \theta_{1,2}} \, A^{\mu \alpha \beta}_E q_\mu \, U_\alpha U_\beta \, .
\end{align*}
This is an identity for eqs. \eqref{2c} (see also the second equation after \eqref{1c}). \\
We now proceed evaluating the other components of $T^{\alpha \beta} \, - \, T^{\alpha \beta}_E$. We use \eqref{1c}$_1$ contracted with $h_\alpha^\theta U_\beta \, c \, p^\alpha p^\beta \left( 1+\frac{\mathcal{I}}{m c^2} \right) \, \varphi (\mathcal{I})$ and integrate in $d \, \mathcal{I} \, d \, \vec{P}$ to find 
\begin{align}\label{5c}
&\left( T^{\alpha \beta} \, - \, T^{\alpha \beta}_E \right) h_\alpha^\theta U_\beta = 
\frac{m \, c^2 \tau}{k_B} \left( h_\alpha^\theta T^{\alpha \delta}_E   \, \partial_\delta \lambda^E +  h_\alpha^\theta A^{\alpha \delta \nu}_E   \, \partial_\delta \lambda^E_\nu \right) - \, \frac{3}{c^4 \rho \, \theta_{1,2}} \, A^{\mu \alpha \beta}_E q_\mu \, h_\alpha^\theta U_\beta  = \nonumber\\ 
&= - \, \frac{m \, c^2 \tau}{k_B} \left( p \, h^{\theta \delta}_E   \, \partial_\delta \lambda^E +  \frac{2}{3} \, \rho \, c^2 \, \theta_{1,2} \, h^{\theta ( \delta} U^{\nu )} \, \partial_\delta \lambda^E_\nu \right) - \,  q^{\theta}  =  - \,  q^{\theta} \, ,
\end{align}
where, in the last passage, we have used \eqref{3c}. The result is an identity. We note that, in the 6 moments model, the left hand side of \eqref{5c} is zero, so that the right hand side is $- \,  q^{\theta}$ must be zero; but we have said, after eq. \eqref{4c} that in this case the physical requirement $V^\alpha - V^\alpha_E=0$ would be violated. This means that this approach cannot be applied to the case of 6 moments. This is not surprising because it has been shown in eq. (19) of \cite{{Gera}} (see also \cite{1a}) that  the optimal choices of moments are  $N=0$ (trivial case with only the conseration law of mass), $N=1$ (only the 5 Euler's Equations where there is no production term), $N=2$ (the 15 moments model), $N=3$ (the 35 moments model) and so on. The 6 moments model is not present in this hierarchy, but it can be considered a subsystem of the 15 moments model by putting $q^{\theta}=0$ (forgetting the role it played in building the model and simply eliminating eq. \eqref{4c}). From this perspective the article \cite{OMS} can be considered correct. \\
Finally, we multiply eq. \eqref{1c}$_1$ by $h_\alpha^\theta \, h_\beta^\psi \, c p^\alpha p^\beta \left( 1+\frac{\mathcal{I}}{m c^2} \right) \, \varphi (\mathcal{I})$ and integrate in $d \, \mathcal{I} \, d \, \vec{P}$; so we obtain
\begin{align*}
h_\alpha^\theta \, h_\beta^\psi \left( T^{\alpha \beta} \, - \, T^{\alpha \beta}_E \right)  = h_\alpha^\theta \, h_\beta^\psi \left[ \frac{m \,\tau}{k_B} \, \left( A^{* \delta \alpha \beta}  \partial_\delta \, \lambda^E  + 
\, A^{* \delta \alpha \beta \nu}  \, \partial_\delta \, \lambda^E_\nu \right)  \, - 
 \frac{3}{c^4 \, \rho \, \theta_{1,2}} \, q_\mu \, A^{\mu \alpha \beta}_E \right] = \\
 =  \frac{m \,\tau}{k_B} \left[
\frac{1}{3} \, \rho \, c^2 \, \theta_{1,2}^* \, h^{\theta \psi} U^\delta \partial_\delta \, \lambda^E  + \left( \frac{1}{6} \, \rho \, c^2 \, \theta_{1,3}^* \, h^{\theta \psi} \, U^\delta \, U^\nu \, + \, \rho \, c^4 \, \theta_{2,3}^* \, h^{( \theta \psi} \, h^{\delta \nu )}\right)  \partial_\delta \, \lambda^E_\nu \right] \, .
\end{align*}
This equation, contracted with $h_{\theta \psi}$ gives
\begin{align}\label{6c}
&\Pi = \frac{m \,\tau}{k_B} \left[ \frac{1}{3} \,
\rho \, c^2 \, \theta_{1,2}^* \,  U^\delta \partial_\delta \, \lambda^E  + \left( \frac{1}{6} \, \rho \, c^2 \, \theta_{1,3}^*  \, U^\delta \, U^\nu \, + \, \frac{5}{9} \, \rho \, c^4 \, \theta_{2,3}^*  \, h^{\delta \nu} \right)  \partial_\delta \, \lambda^E_\nu \right] = - \, \nu \, \partial_{\alpha} \, U^\alpha \, , \\
&\mbox{with} \quad \nu =\nonumber\\& - \, \frac{m \,\tau}{k_B} \, \left[ \left| \begin{matrix}
\rho && \frac{e}{c^2} \\
&& \\
\frac{e}{c^2} && \rho \, \theta_{0,2}
\end{matrix} \right|^{-1} \, \left(\frac{1}{3} \, \rho \, c^2 \, \theta_{1,2}^*  \, 
\left| \begin{matrix}
p && \frac{e}{c^2} \\
&& \\
\frac{1}{3} \, \rho \, c^2 \, \theta_{1,2} && \rho \, \theta_{0,2}
\end{matrix} \right| 
 \,  +\right. \right.\nonumber\\ &\left.\left. + \frac{1}{6} \, \rho \, c^2 \, \theta_{1,3}^*  \,  \, \left| \begin{matrix}
\rho && p \\
&& \\
\frac{e}{c^2} &&  \frac{1}{3} \, \rho \, c^2 \, \theta_{1,2}
\end{matrix} \right| \right) \,  
-  \, \frac{5}{9} \, \rho \, c^4 \, \theta_{2,3}^*  \right] 
\, ,
\end{align}
where eqs. \eqref{2c} have been used. Moreover, contracting equation \eqref{5c} with $h_\theta^{< \gamma} \, h_\psi^{\phi >} = h_\theta^\gamma \, h_\psi^\phi \, - \, \frac{1}{3} \, h_{\theta \psi} h^{\gamma \phi}$ it gives us
\begin{align}\label{7c}
t_{< \gamma \phi >}  = 2 \, \mu \, h^\alpha_{< \beta} \, h^\mu_{\gamma >} \, \partial_{\alpha} \, U_\mu \, , \quad \mbox{with} \quad \mu= 
\frac{1}{3} \, \frac{m \,\tau}{k_B T} \,
 \rho \, c^4 \, \theta_{2,3}^*  \, .
\end{align}
The equations \eqref{4c}, \eqref{6c} and \eqref{7c}$_1$  are those of Relativistic Ordinary Thermodynamics. \\
In conclusion, with this approach we have obtained the equations of Relativistic Ordinary Thermodynamics with  heat conductivity, bulk viscosity and shear viscosity given respectively by \eqref{4c}$_2$, \eqref{6c}$_2$ and \eqref{7c}$_2$. \newline
It is evident from these expressions that they do not depend on the number of moments of the extended model from which they are derivated.

\section{The non relativistic approach}\label{sec:NR}
In this case the balance equation found in eq. (19)-(20) of \cite{1a} are
\begin{align}\label{1A}
& \partial_t \, H_s^{i_1 \cdots i_h} + \partial_k  \, H_s^{ki_1 \cdots i_h} =  J_s^{i_1 \cdots i_h} \,  \, \mbox{with} \quad s = 0, \cdots , N, \mbox{and} \quad h = 0, \cdots , N-s. 
\end{align} 
In particular, $H_0 = \rho$ is the mass density, $H_0^{i_1}= \rho \, v^{i_1}$ where $v^{i_1}$ is the velocity and \\
$H_1 = 2 \, \rho \, \epsilon \, + \, \rho \, v^{2}$ where $\epsilon$ is the energy density. 
All the variables are expressed in integral form as 
\begin{align}\label{1B}
& H_s^{i_1 \cdots i_h} = m \, \int_{\mathbb{R}^{3}} \int_{0}^{+ \, \infty} f \, \xi^{i_1} \, \cdots \, \xi^{i_h} \, \left( \frac{2 \, \mathcal{I}}{m} \, + \, \xi^2 \right)^s \, \varphi ( \mathcal{I} ) \, d \, \mathcal{I} \, d \, \vec{\xi} \, . 
\end{align} 
The expression of $H_s^{k i_1 \cdots i_h}$ is the same of equation \eqref{1B} but with a further factor $\xi^k$ inside the integral; the expression of $J_s^{i_1 \cdots i_h}$ is the same of \eqref{1B} but with the production density $Q= - \, \frac{f-f_E}{\tau}$ instead of the distribution function $f$. This distribution function has the form 
\begin{align}\label{1C}
& f= e^{-1 - \, \frac{m}{k_B} \, \chi} \, , \quad \chi = \sum_{h=0}^{N} \sum_{s=0}^{N-h}
\lambda^s_{i_1 \cdots i_h} \xi^{i_1} \, \cdots \, \xi^{i_h} \, \left( \frac{2 \, \mathcal{I}}{m} \, + \, \xi^2 \right)^s \, .
\end{align} 
We prove now that the Chapman-Enskog method and the Maxwellian Iteration method give the same result for polyatomic gases and with whatever number of moments. This was already proved in \cite{1b} but only for monoatomic gases with 14 moments. \newline
Let us start with the  Chapman-Enskog Method where the Boltzamnn equation and the conservation laws of mass, momentum and energy are considered: 
\begin{align}\label{1D}
& \partial_t \, f + \xi^k \partial_k  \, f =  - \, \frac{f-f_E}{\tau} \, , \, \partial_t \, H_0 + \partial_k  \, H_0^{k} = 0 \, , \, \partial_t \, H_0^{i} + \partial_k  \, H_0^{ki} = 0 \, , \, \partial_t \, H_1 + \partial_k  \, H_1^{k} =  0 \, . 
\end{align} 
After that, the following steps are followed:
\begin{enumerate}
	\item The left hand sides of \eqref{1D} are calculated at equilibrium, while the right hand sides at first order with respect to equilibrium
\begin{align}\label{1E}
\begin{split}
&  \partial_t \, f_E + \xi^k \partial_k  \, f_E  =   \frac{f-f_E}{\tau} \, , \\
&  \partial_t \, H_0 + \partial_k  \, H_0^{k} = 0 \, , \, \partial_t \, H_0^{i} + \partial_k  \, H_{0E}^{ki} = 0 \, , \, \partial_t \, H_1 + \partial_k  \, H_{1E}^{k} =  0 \, . 
\end{split}
	\end{align} 	
\item The derivatives with respect to time of the independent variables $\rho$, $v^i$, $T$ are obtained from \eqref{1E}$_{2-4}$ and substituted in \eqref{1E}$_{1}$ which, after that, depends only on the independent variables and on their derivatives with respect to $x^k$. 
\item The new eq. \eqref{1E}$_{1}$ is multiplied by  $m \, \xi^{i_1} \,  \xi^{i_2} \, \varphi ( \mathcal{I} )$ and integrated with respect to $d \, \mathcal{I} \, d \, \vec{\xi}$. 
\newline
\vspace{0.4cm}
\begin{picture}(100,70)
\put(2.3,50){$\overline{\partial_t \, H_{0E}^{i_1i_2}} + \partial_k  \, H_{0E}^{k i_1i_2} = - \, \left( H_0^{i_1i_2} \, - \, H_{0E}^{i_1i_2} \right)\, \frac{1}{\tau}$}
\put(43,30){\vector(-3,2){20}} \put(117,30){\vector(3,2){20}}
\put(37,10){same variable}
\end{picture}
\newline
%\vspace{0.4cm}
As a result of the above figure we get the Navier-Stokes equations with a precise expression of the bulk viscosity and of the shear viscosity (in the above figure the overline denotes the fact that, after derivating with respect to time, the above found time derivatives of the independent variables have been substituted).
\newline
Similarly, multiplication of the new eq. \eqref{1E}$_{1}$  by  $m \, \xi^{i_1} \,  \left( \frac{2 \, \mathcal{I}}{m} \, + \, \xi^2 \right)^s  \, \varphi ( \mathcal{I} )$ and integration with respect to $d \, \mathcal{I} \, d \, \vec{\xi}$ gives 
\newline
\vspace{0.4cm}
\begin{picture}(100,70)
\put(2.3,50){$\overline{\partial_t \, H_{1E}^{i_1}} + \partial_k  \, H_{1E}^{k i_1}  = - \, \left( H_1^{i_1} \, - \, H_{1E}^{i_1} \right)\, \frac{1}{\tau} $}
\put(43,30){\vector(-3,2){20}} \put(105,30){\vector(3,2){20}}
\put(37,10){same variable}
\end{picture}
\newline
%\vspace{0.4cm}
from which the Fourier equation with a precise expression of the heat conductivity. The expressions found for the bulk viscosity, the shear viscosity and the heat conductivity do not depend on the number $N$ because they use equation which are present in every model with $N \geq 1$. 
\end{enumerate}
Now, mathematically speaking, nothing changes if we swap the order of the steps 2 and 3;
but in this way we obtain the steps of the Maxwellian Iteration. So we can say that, in the non relativistic case, the Chapman-Enskog Method and the Maxwellian Iteration give the same result. \\
So it is natural to ask why in the relativistic case the two methods give different results. To understand it, let us repeat the same steps in the relativistic case. \\
With the  {\bf Chapman-Enskog Method} the Boltzamnn equation and the conservation laws of mass, momentum-energy, with the left hand sides calculated at equilibrium, are considered: 
\begin{align}\label{1F}
\begin{split}
& p^\alpha \, \partial_\alpha \, f_E = \frac{U_\mu}{c^2 \tau}\left[\left(f_E-f\right) \, p^\mu-f_E \,  q_\gamma \, \frac{3}{m \, c^4 \rho \, \theta_{1,2}} \, p^\mu p^\gamma \, \left( 1+\frac{\mathcal{I}}{m c^2}\right) \right] \, . \\
& \partial_\alpha V^{\alpha} = 0 \, , \quad
\partial_\alpha T^{\alpha \beta} = 0 \, , 
\end{split}
\end{align}
After that, the following steps are followed:
\begin{enumerate}
	\item The deviation of the distribution function from its value at equilibrium is calculated in terms of $\partial_\alpha \lambda_E$ and  $\partial_\alpha \lambda_E^\mu$
from \eqref{1F}$_1$  and used in the definition of $T^{\alpha \beta}$ which now becomes
\begin{align}\label{1G}
 T^{\alpha \beta} -T^{\alpha \beta}_E = - c^3 \tau \, \int_{\mathbb{R}^{3}} \int_{0}^{+ \infty} \frac{p^\alpha \partial_\alpha f_E}{p^\mu U_\mu}  p^\alpha p^\beta \left( 1+\frac{\mathcal{I}}{m c^2}\right) \, \varphi ( \mathcal{I} ) \, d \, \mathcal{I} \, d \, \vec{p} 
- 3 \frac{q_\gamma}{c^4 \theta_{1,2}} A^{\gamma \alpha \beta}_E \, , 
\end{align}
\item The quantities $\partial_\alpha \lambda_E$ and $U^\alpha U^\mu \partial_\alpha \lambda_E^\mu$
 are calculated from \eqref{1F}$_{2,3}$ and substituted in \eqref{1G}. From the resulting expression, the bulk viscosity, the shear viscosity and the heat conductivity can be obtained and they do not depend on the number $N$ because they use equation which are present in every model with $N \geq 1$. 
\end{enumerate}
There is also the oppotunity to modify a little the procedure, by taking the last term in \eqref{1F}$_1$ to the left hand side before calculating the left hand sides at equilibrium; in this case it will disappear and, consequently, also the last term in \eqref{1G} will be no more present. \\
Instead of this, with the  {\bf Maxwellian Iteration}  
\begin{enumerate}
	\item  The conservation laws of mass, momentum-energy, and the balance equation for the triple tensor with the left hand sides calculated at equilibrium, are considered:
\begin{align}\label{1H}
\hspace{-7.6cm}\partial_\alpha V^{\alpha}_E = 0 \, , \quad
\partial_\alpha T^{\alpha \beta}_E = 0 \, ,
\end{align}
	\begin{picture}(100,70)
\put(2.3,50){$\partial_\alpha \, A_{E}^{\alpha \beta \gamma} = -  \, \left(  A^{\mu \beta \gamma} \, - \, A_{E}^{\mu \beta \gamma} \right)\, \frac{U_\mu}{c^2 \tau} \, - \,  U_\mu q_\delta A_{E}^{\mu \delta \beta \gamma} \, \frac{3}{\tau \, c^6 \rho \, \theta_{1,2}} \, .$}
\put(43,30){\vector(-3,2){20}} \put(87,30){\vector(3,2){20}} \put(243,30){\vector(-3,2){20}}
\put(7,10){Same variable but different from $T^{\beta \gamma}$.} \put(220,10){ Variable different from $T_E^{\beta \gamma}$ and from $A_{E}^{\mu \beta \gamma}$.}
\end{picture}
\item Some derivatives  of the independent variables  are obtained from \eqref{1H}$_{1,2}$ and substituted in \eqref{1H}$_{3}$. 
	\item The new eq. \eqref{1H}$_{3}$ is used to obtain $\Pi$, $q^\alpha$, $t^{< \beta \gamma >}$ and, consequently,  the bulk viscosity, the shear viscosity and the heat conductivity. This fact could  give rise to some doubts because these coefficients should be obtained from  $T^{\beta \gamma} \, - \, T_{E}^{\beta \gamma}$, not from  $U_\mu \left(A^{\mu \beta \gamma} \, - \, A_{E}^{\mu \beta \gamma} \right)$.  Moreover,  $U_\mu \left(A^{\mu \beta \gamma} \, - \, A_{E}^{\mu \beta \gamma} \right)$ depends not only on $\Pi$, $q^\alpha$, $t^{< \beta \gamma >}$ but also on other variables whose number  increases for increasing values of $N$. So $\Pi$, $q^\alpha$, $t^{< \beta \gamma >}$ must be isolated from the other variables and this means solving some algebraic linear systems depending on $N$. It is therefore not surprising that the solution also depends on $N$. Obviously, this is consequence  of the form of the production term in the right hand side of \eqref{1F}. It remains open the problem to find another expression which respect the requirements of zero production of mass and of momentum-energy, and whose consequent Maxwellian Iteration does not depend on $N$. We can say that another possible expression is 
\begin{align}\label{1I}
Q= - \, \frac{U^\alpha_L \, p^\alpha}{c^2 \tau} \, \left( f-f_E \right) \, ,
\end{align}
where $U^\alpha_L$ is the 4-velocity in the Landau-Liftschiz frame as reported in \cite{1c}, \cite{1d}. But in \cite{1e} it was proved that, up to first order with respect to equilibrium, the expression \eqref{1I} is equivalent to the right hand side of the present eq. \eqref{1F}. So nothing changes by adopting the production term \eqref{1I}. 
\end{enumerate}
In any case, the two procedures have to give the same result at the non relativistic limit. In fact, from $U^\alpha U_\alpha=c^2$, $p^\alpha p_\alpha=m^2c^2$ we have the following decompositions:
\begin{align}\label{1L}
U^\alpha = \Gamma (v) \left( c \, , v^i \right) \, , \quad p^\alpha = m \, \Gamma \left( \frac{p}{m} \right) \left( c \, , \frac{p^i}{m} \right) \, , \quad \mbox{with} \quad \Gamma (v)  = \left( 1- \, \frac{v^2}{c^2} \right)^{-1/2} \, . 
\end{align}
Consequently, the limit for $c \, \rightarrow \, + \infty$ of \eqref{1F} is 
\begin{align}\label{1M}
m \, \left( \partial_t \, f + \xi^k \partial_k  \, f \right) = - \frac{m}{\tau} \, \left( f-f_E \right) \, ,
\end{align}
as in eq. \eqref{1D}$_1$. It follows that both the results of the Chapman-Enskog Method and the Maxwellian Iteration have the same non relativistic limit. In the next section we compute the non relativistic limit.

\subsection{The non relativistic limit of $\chi$, $\mu$ and $\nu$}\label{sec:NR_1}
In this section we prove the convergence in the non relativistic limit of the heat conductivity $\chi$, the shear viscosity $\mu$, and the bulk viscosity $\nu$.

In the previous sections we have introduced the new variables $\theta_{1,1}^*$, $\theta_{1,2}^*$, $\theta_{1,3}^*$, $\theta_{2,3}^*$ which have not studied so far in literature. In order to compute the non relativistic limit of $\chi$, $\mu$ and $\nu$ it is necessary to analyze the non relativistic limit of these new quantitities.
\newline
Taking into account \eqref{A3}, we have 
\begin{align*}
\theta_{1,2}^* = 3 \, \theta_{1,1} \, ,  \, \theta_{1,3}^* = 2 \, \theta_{1,2}  \, .
\end{align*}
So we need only the non relativistic limit of $\theta_{1,1}^*$, $\theta_{2,3}^*$ given by \eqref{A5}. To evaluate them, let us consider the expression of $J_{4,-1}$, i.e.,
\begin{align*}
\begin{split}
& J_{4,-1} =\int_0^{+\infty} e^{- \gamma \, \cosh \, s} \frac{\sinh^4 \, s}{\cosh \, s} \, d \, s = \int_0^{+\infty} e^{- x} \, e^{- \gamma} \frac{\left[ \left( \frac{x}{\gamma} + 1 \right)^2 - \, 1 \right]^{\frac{3}{2}} }{\frac{x}{\gamma} + 1} \, \frac{d \, x}{\gamma} = \\
& = \frac{e^{- \gamma}}{\gamma} \int_0^{+\infty} e^{- x} \, \frac{\left( \frac{x}{\gamma} + 2 \right)^{\frac{3}{2}} }{\frac{x}{\gamma} + 1} \, \left( \frac{x}{\gamma} \right)^{\frac{3}{2}} \, d \, x \,  , 
\end{split}
\end{align*}
where in the first passage we have changed the integration variable according to the law \\
$\cosh \, s = \frac{x}{\gamma} + 1 \quad \rightarrow \quad \sinh \, s \, d \, s= \frac{d \, x}{\gamma}$. Now the Mac-Laurin expansion of the function $g(y)= \frac{\left( y + 2 \right)^{\frac{3}{2}} }{y+ 1}$ around $y=0$ is 
\begin{align*}
g(y) &= 2 \, \sqrt{2} \left( 1 \, - \frac{1}{4} \, y \right) \, + \, y^2 (\cdots)  \, \,  \rightarrow \, \,  \gamma^{\frac{5}{2}} e^\gamma \, J_{4,-1} = \\&= 2 \, \sqrt{2} \, \int_0^{+\infty} e^{- x} \, \left( 1 \, - \frac{1}{4} \, \frac{x}{\gamma}  \, + \, \frac{1}{\gamma^2} (\cdots)  \right) \,  x^{\frac{3}{2}} \, d \, x  = \\&=
 2 \, \sqrt{2} \, \left[ \Gamma\left( \frac{5}{2}\right) \, - \, \frac{1}{4}  \Gamma\left( \frac{7}{2}\right) \frac{1}{\gamma} \, + \, \frac{1}{\gamma^2} (\cdots) \right]=\\& = 2 \, \sqrt{2 \, \pi} \, \left[  \frac{3}{4} \, - \, \frac{15}{32}  \,  \frac{1}{\gamma} \, + \, \frac{1}{\gamma^2} (\cdots) \right]  \, . 
\end{align*}
where in the  last passage we have used the Gamma function 
\begin{align*}
\Gamma (s) = \int_0^{+\infty} e^{- x} \, x^{s-1} \, d \, x 
\end{align*}
defined for $s > 0$ and satisfying the relations $\Gamma (s+1) = s \, \Gamma (s)$,  $\, \, \Gamma\left( \frac{1}{2}\right) = \sqrt{\pi}$. In a similar way we can obtain the expansion of $J_{2,1}$ or we can read it on page 21 of \cite{P-R} and it is 
\begin{align*}
J_{2,1} =  2 \, \sqrt{2 \, \pi} \, e^{- \gamma} \, \gamma^{- 1/2} \, \left[ \frac{1}{4 \, \gamma} \, + \, \frac{15}{32} \, \frac{1}{\gamma^2} \, +  \, \frac{105}{512} \, \frac{1}{\gamma^3} \, - \, \frac{315}{32 \, \cdot \, 128} \, \frac{1}{\gamma^4} \, + \, \cdots \right] \, . 
\end{align*}
It follows that 
\begin{align*}
\frac{\int_{0}^{+ \infty} J_{4,-1}^* \,  \varphi(\mathcal{I}) \, d \, \mathcal{I} }{\int_{0}^{+ \infty} J_{2,1}^* \,  \varphi(\mathcal{I}) \, d \, \mathcal{I}}&= 
\frac{\int_{0}^{+ \infty} \gamma^{\frac{5}{2}} e^{\gamma} \,   J_{4,-1}^* \,  \varphi(\mathcal{I}) \, d \, \mathcal{I} }{\gamma \, \int_{0}^{+ \infty} \gamma^{\frac{3}{2}} e^{\gamma} \,  J_{2,1}^* \,  \varphi(\mathcal{I}) \, d \, \mathcal{I}} = \frac{\int_{0}^{+ \infty} \left( \frac{\gamma}{\gamma^*} \right)^{\frac{5}{2}} e^{\gamma - \gamma^*} \, e^{\gamma^*} \, \gamma^{* \, \frac{5}{2}} \,   J_{4,-1}^* \,  \varphi(\mathcal{I}) \, d \, \mathcal{I} }{\gamma \, \int_{0}^{+ \infty} \left( \frac{\gamma}{\gamma^*} \right)^{\frac{3}{2}} e^{\gamma - \gamma^*} \, e^{\gamma^*} \, \gamma^{* \, \frac{3}{2}} \,  J_{2,1}^* \,  \varphi(\mathcal{I}) \, d \, \mathcal{I}} = \\
&=  \frac{\int_{0}^{+ \infty} \left( \frac{\gamma}{\gamma^*} \right)^{\frac{5}{2}} e^{\gamma - \gamma^*} \, 2 \, \sqrt{2 \, \pi} \, \left[  \frac{3}{4} \, - \, \frac{15}{32}  \,  \frac{1}{\gamma^*} \, + \, \frac{1}{\gamma^{*2}} (\cdots) \right]  \,  \varphi(\mathcal{I}) \, d \, \mathcal{I} }{\gamma \, \int_{0}^{+ \infty} \left( \frac{\gamma}{\gamma^*} \right)^{\frac{3}{2}} e^{\gamma - \gamma^*} \,\, 2 \, \sqrt{2 \, \pi} \, \left[  \frac{1}{4} \, + \, \frac{15}{32}  \,  \frac{1}{\gamma^*} \, + \, \frac{1}{\gamma^{*2}} (\cdots) \right] \,  \varphi(\mathcal{I}) \, d \, \mathcal{I}} \, , 
\end{align*}
and, consequently,
\begin{align*}
&\gamma \, \left( \gamma \, \frac{\int_{0}^{+ \infty} J_{4,-1}^* \,  \varphi(\mathcal{I}) \, d \, \mathcal{I} }{\int_{0}^{+ \infty} J_{2,1}^* \,  \varphi(\mathcal{I}) \, d \, \mathcal{I}} \, - \, 3 \right) = \\
&= \frac{\int_{0}^{+ \infty} \left( \frac{\gamma}{\gamma^*} \right)^{\frac{3}{2}} e^{\gamma - \gamma^*} \, \left[  \frac{3}{4} \gamma \, \left( \frac{\gamma}{\gamma^*} \, - \, 1 \right)\, -  \, \frac{15}{32}  \,  \left( \frac{\gamma}{\gamma^*} \right)^2   \, - \, \frac{45}{32}  \,  \frac{\gamma}{\gamma^*} \, + \, \frac{1}{\gamma^{*}} (\cdots)\right]  \,  \varphi(\mathcal{I}) \, d \, \mathcal{I} }{\int_{0}^{+ \infty} \left( \frac{\gamma}{\gamma^*} \right)^{\frac{3}{2}} e^{\gamma - \gamma^*}  \, \left[  \frac{1}{4} \, + \,  \frac{1}{\gamma^*}  (\cdots) \right] \,  \varphi(\mathcal{I}) \, d \, \mathcal{I}} \, .
\end{align*}
Since we have 
\begin{align*}
e^{\gamma - \gamma^*}= e^{- \frac{\mathcal{I}}{k_B T}} \, , \, \gamma \, \left( \frac{\gamma}{\gamma^*} \, - \, 1 \right) = \frac{\gamma}{\gamma^*} \left( \gamma - \gamma^* \right) = - \, \frac{\gamma}{\gamma^*} \, \frac{\mathcal{I}}{k_B T} \, , 
\end{align*}
we can write
\begin{align*}
\lim_{\gamma \, \rightarrow \, + \infty} \, \gamma \, \left( \gamma \, \frac{\int_{0}^{+ \infty} J_{4,-1}^* \,  \varphi(\mathcal{I}) \, d \, \mathcal{I} }{\int_{0}^{+ \infty} J_{2,1}^* \,  \varphi(\mathcal{I}) \, d \, \mathcal{I}} \, - \, 3 \right) 
=  - \, 3 \, \frac{\int_{0}^{+ \infty}  e^{- \, \frac{\mathcal{I}}{k_B T}} \, \frac{\mathcal{I}}{k_B T}    \,  \varphi(\mathcal{I}) \, d \, \mathcal{I} }{\int_{0}^{+ \infty}  e^{- \, \frac{\mathcal{I}}{k_B T}}  \,  \varphi(\mathcal{I}) \, d \, \mathcal{I}}
- \,  \frac{15}{2} \,  .
\end{align*}
Consequently, we get 
\begin{align*}
\frac{\int_{0}^{+ \infty} J_{4,-1}^* \,  \varphi(\mathcal{I}) \, d \, \mathcal{I} }{\int_{0}^{+ \infty} J_{2,1}^* \,  \varphi(\mathcal{I}) \, d \, \mathcal{I}}   
=  3 \, \frac{1}{\gamma} \, + \, \left( - \, 3 \, \frac{\int_{0}^{+ \infty}  e^{- \, \frac{\mathcal{I}}{k_B T}} \, \frac{\mathcal{I}}{k_B T}    \,  \varphi(\mathcal{I}) \, d \, \mathcal{I} }{\int_{0}^{+ \infty}  e^{- \, \frac{\mathcal{I}}{k_B T}}  \,  \varphi(\mathcal{I}) \, d \, \mathcal{I}}
- \,  \frac{15}{2} \right) \frac{1}{\gamma^2} \, + \, \frac{1}{\gamma^3}  (\cdots) \,  , 
\end{align*}
\begin{align*}
\theta_{1,1}^* =  \frac{1}{\gamma} \, + \, \left( - \, \frac{\int_{0}^{+ \infty}  e^{- \, \frac{\mathcal{I}}{k_B T}} \, \frac{\mathcal{I}}{k_B T}    \,  \varphi(\mathcal{I}) \, d \, \mathcal{I} }{\int_{0}^{+ \infty}  e^{- \, \frac{\mathcal{I}}{k_B T}}  \,  \varphi(\mathcal{I}) \, d \, \mathcal{I}}
- \,  \frac{5}{2} \right) \frac{1}{\gamma^2} \, + \, \frac{1}{\gamma^3}  (\cdots) = \\
= \frac{p}{\rho \, c^2} \, + \, \left( - \, \frac{\int_{0}^{+ \infty}  e^{- \, \frac{\mathcal{I}}{k_B T}} \, \frac{\mathcal{I}}{k_B T}    \,  \varphi(\mathcal{I}) \, d \, \mathcal{I} }{\int_{0}^{+ \infty}  e^{- \, \frac{\mathcal{I}}{k_B T}}  \,  \varphi(\mathcal{I}) \, d \, \mathcal{I}}
- \,  \frac{5}{2} \right) \frac{1}{\gamma^2} \, + \, \frac{1}{\gamma^3}  (\cdots) \,  .
\end{align*}
We can apply this result in \eqref{4c}$_2$, jointly with $\theta_{1,2}^* = 3 \,\theta_{1,1} $  , and have that the heat conductivity has the form
\begin{align*}
&\chi = - \frac{m \, \tau c^8}{9 \, k_B T^2} \, \frac{\rho^2}{p} \, \theta_{1,2}\\& \left[ 3 \, \theta_{1,1} \,  -  \, \theta_{1,2}  + \, \frac{\rho \, c^2}{p} \, \theta_{1,2} \left( \frac{\int_{0}^{+ \infty}  e^{- \, \frac{\mathcal{I}}{k_B T}} \, \frac{\mathcal{I}}{k_B T}    \,  \varphi(\mathcal{I}) \, d \, \mathcal{I} }{\int_{0}^{+ \infty}  e^{- \, \frac{\mathcal{I}}{k_B T}}  \,  \varphi(\mathcal{I}) \, d \, \mathcal{I}}
+ \,  \frac{5}{2} \right) \frac{1}{\gamma^2} -\frac{\rho \, c^2}{p} \, \theta_{1,2} \frac{1}{\gamma^3}  (\cdots) \right] \,  . 
\end{align*}
Moreover, from eq. (11) of \cite{Arx}, we have that the exact expressions 
\begin{align*}
\theta_{1,1} = \frac{p}{\rho} \, \frac{1}{c^2} \, , \, \theta_{1,2} = 3 \, \frac{p}{\rho} \, \frac{1}{c^2} \, + \,  3 \, \frac{p}{\rho} \, g_1 \, \frac{1}{c^4} \quad \mbox{with} \quad g_1= \frac{e- \rho c^2 \, + \, p}{\rho} \, , 
\end{align*}
so that we find 
\begin{align*}
\chi = - \, \frac{m \, \tau c^2}{k_B T^2} \, \rho  \,  \left[ -   \, \frac{p}{\rho} \, g_1 \,   +   \, \left( \frac{\int_{0}^{+ \infty}  e^{- \, \frac{\mathcal{I}}{k_B T}} \, \frac{\mathcal{I}}{k_B T}    \,  \varphi(\mathcal{I}) \, d \, \mathcal{I} }{\int_{0}^{+ \infty}  e^{- \, \frac{\mathcal{I}}{k_B T}}  \,  \varphi(\mathcal{I}) \, d \, \mathcal{I}}
+ \,  \frac{5}{2} \right) \left( \frac{p}{\rho} \right)^2   \, -  \,   \frac{p^3}{\rho^3 c^2}  (\cdots) \right] \,  . 
\end{align*}
By performing similar calculations, we find that
\begin{align*}
g_1 = \left( \frac{\int_{0}^{+ \infty}  e^{- \, \frac{\mathcal{I}}{k_B T}} \, \frac{\mathcal{I}}{k_B T}    \,  \varphi(\mathcal{I}) \, d \, \mathcal{I} }{\int_{0}^{+ \infty}  e^{- \, \frac{\mathcal{I}}{k_B T}}  \,  \varphi(\mathcal{I}) \, d \, \mathcal{I}}
+ \,  \frac{5}{2} \right) \,\frac{p}{\rho} + \, \frac{1}{\gamma} \, (\cdots) \,  , 
\end{align*}
so that the above expression of the heat conductivity $\chi $ has a finite non relativistic limit. \\
To evaluate the non relativistic limit of the shear viscosity $\mu$ in \eqref{7c}$_2$, we need the expression of $\theta_{2,3}^*$ and of $J_{6,-1}$. With similar computations we obtain that 
$c^4 \, \theta_{2,3}^*$ and $\mu$ have a finite limit
\begin{align*}
\lim_{c \, \rightarrow \, + \infty} c^4 \, \theta_{2,3}^* = 3 \, \left( \frac{p}{\rho} \right)^2 \quad \mbox{and} \quad  \lim_{c \, \rightarrow \, + \infty} \mu= 
\, \tau \, p  \, . 
\end{align*}
Moreover, with similar computations we obtain that $\nu$ is convergent in the non relativistic limit.

\section{Summary}\label{sec4}
In this article we have described how it is possible to reconstruct, as a first iteration, the laws of the Relativistic Ordinary Thermodynamics starting from the laws of the 
Relativistic Extended Thermodynamics of polyatomic gases by using two different iteration methods. In literature, two procedures are used which are the so-called Maxwellian Iteration and the Chapman-Enskog Method. Both of these methods lead to the relativistic version of the Navier-Stokes and Fourier laws, i.e, the so-called Eckart equations as a first iteration. It is well known that the relativistic version of the Navier-Stokes and Fourier laws are two fundamental
laws of Relativistic Ordinary Thermodynamics and in these equations the following remarkable physical quantities
appear as coefficients: the heat conductivity $\chi$, the shear viscosity $\mu$, and the bulk viscosity $\nu$. We have proved that the expressions of $\chi$, $\mu$, and $\nu$ obtained via the Chapman-Enskog method do not depend on $N$, whereas these expressions obtained through the Maxwellian Iteration depend on $N$. In order to make clear this difference we describe our main results giving more details.\newline
First of all, we observe that we have found the following expressions for the shear viscosity $\mu$ by using the Maxwellian Iteration method in the case $N=3$ (see eq. \eqref{29}) and in the case $N=2$ (see eq. \eqref{number_sh})
\begin{align*}
\mu &= \frac{5}{3} \, \frac{m \, c^4 \rho \, \tau}{k_B T}  \, 
\frac{\theta_{2,3} }{\theta_{2,4}  }  \, \theta_{2,3}\, ,\, \quad \text{case}\,N=2 \,\,
\\&{}\\
\mu&= - \, \frac{c^4 m \, \rho \, \tau}{2 \, k_B T} \, 
\frac{\left| \begin{matrix}
	\frac{2}{15} \, \theta_{2,4}  &&  \frac{2}{15} \, \theta_{2,5}  && -  \, \frac{2}{3} \,   \theta_{2,3} \\
	&&&& \\
\frac{2}{45} \, \theta_{2,5}  &&  \frac{1}{35} \, \theta_{2,6}  && -  \, \frac{2}{15} \,   \theta_{2,4} \\
	&&&& \\
\frac{2}{3} \, \theta_{2,3}  &&  \frac{2}{5} \, \theta_{2,4}  && 0
	\end{matrix}\right|}{\left| \begin{matrix}
	\frac{2}{15} \, \theta_{2,4}  &&  \frac{2}{15} \, \theta_{2,5}   \\
	&& \\
	\frac{2}{45} \, \theta_{2,5}  &&  \frac{1}{35} \, \theta_{2,6}  
	\end{matrix}\right|} \,,\,\quad \text{case}\,N=3
\end{align*} 
It is immediate to realize that these two expressions of $\mu$ are, in general, different each other and then we can conclude that the Maxwellian Iteration give us in the relativistic case a result depending, in general, on the number of moments $N$. Analogous conclusions can be reached by observing the different expressions of $\chi$ (compare eqs. \eqref{27} and \eqref{31}) and $\nu$ (compare eqs. \eqref{25} and \eqref{30}) when one uses the Maxwellian Iteration method in the relativistic cases for polyatomic gases in the cases $N=3$ and $N=2$, rispectively.

Let us now look at the expressions obtained for $\chi$, $\nu$ and $\mu$ by using the Chapman-Enskog Method in the relativistic case for a polyatomic gas with an arbitrary value of $N$ (which are written below for the convenience of the reader)
\begin{align*}
\nu &= - \frac{m \,\tau}{k_B} \, \left[ \left| \begin{matrix}
\rho && \frac{e}{c^2} \\
&& \\
\frac{e}{c^2} && \rho \, \theta_{0,2}
\end{matrix} 
\right|^{-1} \, \left(  \frac{1}{3} \, \rho \, c^2 \, \theta_{1,2}^*  \, \left| 
\begin{matrix}
p && \frac{e}{c^2} \\
&& \\
\frac{1}{3} \, \rho \, c^2 \, \theta_{1,2} && \rho \, \theta_{0,2}
\end{matrix} \right| 
 \,  + \right. \right. \\
&+ \left.  \, \frac{1}{6} \, \rho \, c^2 \, \theta_{1,3}^*  \,  \, \left| \begin{matrix}
\rho && p \\
&& \\
\frac{e}{c^2} &&  \frac{1}{3} \, \rho \, c^2 \, \theta_{1,2}
\end{matrix} \right| \right) \,  
\left. -  \, \frac{5}{9} \, \rho \, c^4 \, \theta_{2,3}^*  \right]   \, , \\
 \chi &= - \, \frac{m \, \tau c^8}{9 \, k_B T^2} \, \frac{\rho^2}{p} \, \theta_{1,2} \left( \theta_{1,2}^* \,  - \, \frac{\rho \, c^2}{p} \, \theta_{1,2} \theta_{1,1}^* \right) \, ,  \\
 \mu &= \frac{1}{3} \, \frac{m \,\tau}{k_B T} \,
 \rho \, c^4 \, \theta_{2,3}^*   \, ,
\end{align*}
where $\theta_{k,j}$ and $\theta^*_{k,j}$ are introduced in eqs. \eqref{SC} and \eqref{A2}, rispectively.
Since these expressions of $\chi$, $\nu$ and $\mu$ do not depend on $N$ we can conclude that the Chapman-Enskog Method furnishes results which do not depend on the number of the moments $N$ in the relativistic case. The convergence of $\nu, \mu, \chi$ in the non relativistic limit have been proved in section \ref{sec:NR_1}.

Moreover, in section {\ref{sec:NR} it has been proved that the Maxwellian Iteration and the Chapman-Enskog Method lead at the same results in nonrelativistic case.

Finally, we want to conclude this section with an important observation. Of course, if one uses the Maxwellian Iteration Method for polyatomic gas in the relativistic case the results depend on the choice of the production term $Q$ defined in equation \eqref{1}. So a natural problem (still open) is the determination of a specific function $Q$ such that the requirements of zero production of mass and of momentum-energy are satisfied and whose consequent Maxwellian Iteration does not depend on the number of the moments $N$. This remains an open problem.

Another open problem is the limit in the case of the Maxwellian Iteration for $N$ going to infinity. Obviously, if this would be possible, the result will not depend on $N$, but we do not know if this will be the same of Ordinary Thermodynamics obtained with the Chapman-Enskog method being based on the non proved paradigma that``the truncated moment approach is to consider an approximation of the Boltzmann equation that corresponds to infinite moments''. \\ 
In past times it was thought that this was the solution of the problem of the loss of hyperbolicity of the field equations (See \cite{Bri}, \cite{R-T}. But this was not true; in fact, Struchtrup (\cite{Struc1}, \cite{Struc2}) proved that with an increasing number of moments the hyperbolicity region does not increase. In \cite{Bri} a second order approach with respect to equilibrium was tested and the result showed that  the hyperbolicity region increases. Finally, in \cite{Pen}, 
in section 2, the following facts have been proved:
\begin{enumerate}
	\item For every truncated system there exists full hyperbolicity if only the Lagrange multipliers are taken as independent variables;
	\item The hyperbolicity region appears only as a consequence of the approximations (usually at first order with respect to equilibrium) involved in the tranformation from the Lagrange multipliers to physical variables;
	\item The full hyperbolicity is recovered if the above transformation is performed up to whatever order with respect to equilibrium.
\end{enumerate}
So the full hyperbolicity is not achieved by increasing the number of moments (or taking its limit for $N$ going to infinity), but eliminating the approximations in the passage from the Lagrange multipliers to physical variables. 
\newline
Based on this fact, we are not able to anticipate how the model with infinite moments can influence the present problem of the transition to Ordinary Thermodynamics. It needs to be tested.

\section*{Acknowledgments}
The authors would like to thank the two anonymous refeees and C. van der Mee whose suggestions, comments and remarks helped us to improve the quality of the paper.
The authors have been partially supported by INdAM-GNFM. Moreover, one of the authors (FD) has been partially supported by Ministero dell'Universit\`a e della Ricerca of Italy (MIUR) under the PRIN project
``2022TEB52W - The charm of integrability: from nonlinear waves to random matrices''.

\appendix

\section{Some integrals necessary for recovering OT with the Chapman-Enskog Method}
We define 
\begin{align}\label{A0}
 A^{* \alpha_1 \cdots \alpha_{n+1}}= \frac{c^3}{m^{n-2}} \int_{\Re^3} \int_{0}^{+ \infty} \frac{f_E }{U_\mu p^\mu}\, p^{\alpha_1} \cdots p^{\alpha_{n+1}} \left( 1 \, + \, \frac{\mathcal{I}}{m \, c^2} \right)^{n-1} \,  \varphi(\mathcal{I}) \, d \, \mathcal{I} \, d \, \vec{P} \, .
\end{align}
We see that $A^{* \alpha_1 \cdots \alpha_{n+1 } }_E$ is like $A^{\alpha_1 \cdots \alpha_{n+1 } }_E$ but with the function to be integrated which \\
\\
is now divided by $\cosh \, s \left( 1 \, + \, \frac{\mathcal{I}}{m \, c^2} \right)$. 
Consequently, we find the expressions corresponding to  \eqref{AE}, \eqref{SC}, i.e., 
\begin{align}\label{A1}
A^{* \alpha_1 \cdots \alpha_{n+1 } }_E= \sum_{k=0}^{\left[ \frac{n+1}{2} \right]} \rho c^{2k} \theta_{k,n}^* \, h^{( \alpha_1 \alpha_2 } \cdots k^{ \alpha_{2k-1} \alpha_{2k} } U^{\alpha_{2k+1} } \cdots U^{\alpha_{n+1} ) } \, .
\end{align}
where the scalar coefficients $\theta_{k,n}^*$ are 
\begin{align}\label{A2}
\theta_{k,n}^* = \frac{1}{2k+1}  \begin{pmatrix}
n+1 \\ 2k
\end{pmatrix} \frac{\int_0^{+\infty} J_{2k+2,n-2k}^* \, \left( 1 + \frac{\mathcal{I}}{m c^2} \right)^{n-1} \, \phi(\mathcal{I})  \, d \, \mathcal{I}}{\int_0^{+\infty} J_{2,1}^* \,  \phi(\mathcal{I})  \, d \, \mathcal{I}} \, , 
\end{align}
where $J_{m,n}(\gamma)= \int_0^{\infty} e^{-\zg \cosh{s}}\cosh^n{s}\sinh^m{s}\, ds, \,\, \zg=\frac{m c^2}{k_B T}, \,\, J^*_{m,n}= J_{m,n}\left[\zg\left(1+\frac{\mathcal{I}}{mc^2}\right)\right]$.\newline
By comparing this last equation with  \eqref{SC}, we find that 
\begin{align}\label{A3}
\theta_{k,n}^* = \frac{n+1}{n+1 - 2 k} \, \theta_{k,n-1} \, , \quad \mbox{for every $k$ such that} \quad n+1 > 2k \, .
\end{align}
From this last equation it follows that only the expressions for $n+1=2k$  are present (which means that only the case $n$ odd has to be considered), and for these cases, eq. \eqref{A2} gives 
\begin{align}\label{A4}
\theta_{k,2k-1}^* = \frac{1}{2k+1} \, \frac{\int_0^{+\infty} J_{2k+2,-1}^* \, \left( 1 + \frac{\mathcal{I}}{m c^2} \right)^{2k-2} \, \phi(\mathcal{I})  \, d \, \mathcal{I}}{\int_0^{+\infty} J_{2,1}^* \,  \phi(\mathcal{I})  \, d \, \mathcal{I}} \, . 
\end{align}
The expressions are necessary
with $k=1$ and with $k=2$, i.e., 
\begin{align}\label{A5}
 \theta_{1,1}^* = \frac{1}{3} \, \frac{\int_0^{+\infty} J_{4,-1}^* \,  \phi(\mathcal{I})  \, d \, \mathcal{I}}{\int_0^{+\infty} J_{2,1}^* \,  \phi(\mathcal{I})  \, d \, \mathcal{I}} \, , \quad  \theta_{2,3}^* = \frac{1}{5} \, \frac{\int_0^{+\infty} J_{6,-1}^* \, \left( 1 + \frac{\mathcal{I}}{m c^2} \right)^{2} \, \phi(\mathcal{I})  \, d \, \mathcal{I}}{\int_0^{+\infty} J_{2,1}^* \,  \phi(\mathcal{I})  \, d \, \mathcal{I}} \, . 
\end{align}
The derivative of \eqref{A5}$_1$  with respect to $\gamma$ gives
\begin{align*}
\frac{\partial \, \theta_{1,1}^*}{\partial \, \gamma} = \frac{-1}{3} \,  \frac{\int_0^{+\infty} J_{4,0}^* \, \left( 1 + \frac{\mathcal{I}}{m c^2} \right) \,  \phi(\mathcal{I})  \, d \, \mathcal{I}}{\int_0^{+\infty} J_{2,1}^* \,  \phi(\mathcal{I})  \, d \, \mathcal{I}} \, + \, \theta_{1,1}^* \, \frac{e}{\rho \, c^2} = - \, \frac{1}{\gamma}
\, + \, \theta_{1,1}^* \, \frac{e}{\rho \, c^2} \, . 
\end{align*}
From this result it follows
\begin{align}\label{A7}
\frac{e}{\rho \, c^2} =  \frac{1}{\gamma \, \theta_{1,1}^*}
\, + \, \frac{\partial \, }{\partial \, \gamma} \, \ln \theta_{1,1}^* \, . 
\end{align}
Since in literature everything has been expressed in terms of $\frac{e}{\rho \, c^2} $ and its derivatives, we see that now everything is expressed in terms of $\theta_{1,1}^*$, its derivative and of $\theta_{2,3}^*$.

\end{document}